\pdfoutput=1
\documentclass[pdflatex, sn-nature]{sn-jnl}
\usepackage{graphicx}
\usepackage{lipsum}
\usepackage{amsmath,amssymb}%
\usepackage[title]{appendix}%
\usepackage{adjustbox}
\usepackage{booktabs}
\usepackage[stable]{footmisc}
\usepackage[table,xcdraw]{xcolor}
\usepackage{tabularx}
\usepackage{pifont}
\usepackage{listings}
\usepackage[linesnumbered,ruled]{algorithm2e}
\usepackage{multirow}
\usepackage{tikz}
\usepackage{xcolor}
\usepackage{textcomp}%
\usepackage{xparse}
\usepackage{manyfoot}%

\newcommand*\circled[1]{\tikz[baseline=(char.base)]{
            \node[shape=circle,draw,inner sep=0.8pt] (char) {#1};}}
\newcommand{\cmark}{\ding{51}}%
\newcommand{\xmark}{\ding{55}}%

\begin{document}

\title[KubeDSM: A Kubernetes-based Dynamic Scheduling and Migration Framework for Cloud-Assisted Edge Clusters]{KubeDSM: A Kubernetes-based Dynamic Scheduling and Migration Framework for Cloud-Assisted Edge Clusters}

\author[1]{Amirhossein Pashaeehir, Sina Shariati, Shayan Shafaghi, Manni Moghimi, Mahmoud Momtazpour}
\affil*[1]{\orgdiv{Computer Engineering}, \orgname{Amirkabir University of Technology}, \orgaddress{\street{Hafez Ave.}, \city{Tehran}, \state{Tehran}, \country{Iran}}}


\abstract{Edge computing has become critical for enabling latency-sensitive applications, especially when paired with cloud resources to form cloud-assisted edge clusters. However, efficient resource management remains challenging due to edge nodes' limited capacity and unreliable connectivity. This paper introduces KubeDSM, a Kubernetes-based dynamic scheduling and migration framework tailored for cloud-assisted edge environments. KubeDSM addresses the challenges of resource fragmentation, dynamic scheduling, and live migration while ensuring Quality of Service (QoS) for latency-sensitive applications. Unlike Kubernetes' default scheduler, KubeDSM adopts batch scheduling to minimize resource fragmentation and incorporates a live migration mechanism to optimize edge resource utilization. Specifically, KubeDSM facilitates three key operations: intra-edge migration to reduce fragmentation, edge-to-cloud migration during resource shortages, and cloud-to-edge migration when resources become available, thereby increasing the number of pods allocated to the edge. Our results demonstrate that KubeDSM consistently achieves a higher average edge ratio and a lower standard deviation in edge ratios, highlighting its ability to provide more effective and stable scheduling across different deployments. We also explore the impact of migration strategies and Quality of Service (QoS) configurations on the edge ratios achieved by KubeDSM. The findings reveal that enabling migrations significantly enhances the edge ratio by reducing fragmentation. Additionally, KubeDSM's adaptability in respecting QoS requirements while maximizing overall edge ratios is confirmed through different QoS scenarios.}


\maketitle

\section{Introduction}
In recent years, we have witnessed an increasing demand for edge computing environments. Public edge platforms such as Google Distributed Cloud Edge \cite{noauthor_google_nodate}, AWS Local Zones \cite{aws_amazon_nodate} and Azure Public MEC \cite{azure_azure_nodate} has been introduced to enable latency-sensitive workloads to run on edge server clusters and optimize users' quality of experience. Furthermore, the lower latency and higher bandwidth of the 5G technology enables edge-based deployment of many latency-sensitive applications, such as online gaming, remote surgery, real-time video analytics, and edge intelligence. However, there is still a gap for ultra-low-latency applications such as virtual and augmented reality (VR/AR) and autonomous driving (AD)\cite{mohan_pruning_2020}. For example, as characterized by Mohan et al. \cite{mohan_pruning_2020}, AR/VR applications require sub-20ms latency, out of which 13ms should be reserved for display technology, and only around 7ms remains to perform all communications, processings, modellings, and output formation tasks. Next-generation telecommunication technologies like 6G might partially mitigate this challenge. However, efficient resource management techniques still play a significant role in guaranteeing the quality of service (QoS) under such a tight latency constraint. \par
Edge computing systems often comprise heterogeneous nodes with limited computing and storage resources connected through unreliable links. These limitations make resource management, scheduling, and migration particularly challenging in edge-based server clusters. Moreover, due to the stringent latency requirements, resource management techniques must be fast, adding to the complexity of their design. 
Given the limited computing capacity of edge nodes, they may only have enough resources to handle some computational tasks locally. Consequently, to enhance the scalability of edge platforms, several studies have implemented cloud-assisted edge computing, which dynamically allocates resources from the cloud to address the resource shortage at edge sites (\cite{ma_cost-efficient_2021}, \cite{ma_cost-efficient_2017}, \cite{li_heterogeneity-aware_2020}).\par
Furthermore, due to the cost associated with offloading computation from the edge to the cloud, combined with the limited resources of edge servers, optimizing resource utilization on edge nodes is crucial. One way to achieve this is through reducing resource fragmentation by implementing effective migration strategies. By enhancing the utilization of edge resources and ensuring that resources are utilized to their fullest potential, we can significantly reduce the need for computational offloading to the cloud during periods of resource shortage. This not only decreases operational costs but also lowers the average latency of applications running on cloud-assisted edge platforms, thereby improving overall system performance and user experience.\par
This paper introduces a fragmentation-aware, QoS-aware dynamic scheduling and migration framework for cloud-assisted edge server clusters. The framework has been designed and implemented as a live component of Kubernetes and can be directly used in production container orchestration systems such as Kubernetes, K3s and KubeEdge. To the best of our knowledge, this is the first attempt to incorporate a fragmentation-aware scheduling and migration framework for cloud-assisted edge computing into Kubernetes. The main contributions of this work are as follows:
\begin{enumerate}
    \item In contrast to Kubernetes' default scheduler (Kube-scheduler), where pods are scheduled one at a time, the proposed framework handles batch binding of pods to nodes of a Kubernetes cluster to efficiently reduce resource fragmentation
    \item The proposed framework adds a live migration component to Kubernetes to reduce resource fragmentation and efficiently utilize edge resources. The added live component actively monitors the remaining resources on servers of a local edge cluster and tries to:
    \begin{enumerate}
        \item Migrates pods between nodes on the edge cluster to reduce resource fragmentation
        \item Migrates pods from edge to cloud nodes in the event of resource shortage
        \item When enough resources become available, migrate pods from the cloud back to the edge nodes to reduce the applications' average latency. This behaviour, in turn, helps tenant services to maximize their use of available edge resources.
    \end{enumerate}
\end{enumerate}
Experimental results demonstrate that KubeDSM significantly outperforms the default kube-scheduler (K8S) and other baseline configurations. Our comprehensive evaluation reveals that KubeDSM achieves a higher average edge ratio and a lower standard deviation in edge ratios, indicating its superior scheduling efficiency and consistency across various deployments. The impact of migration strategies and QoS configurations further highlights KubeDSM's adaptability and effectiveness in enhancing edge resource utilization.

The rest of the paper is organized as follows. In Section~\ref{section:relatedwork}, we review the related work. Section~\ref{section:systemmodel} defines the system model, detailing the key components and interactions within our proposed architecture. Section~\ref{section:problemdefinition} outlines the problem formulation, where we formulate the problem as a multi-objective Mixed-Integer Linear Programming (MILP) model. In Section~\ref{section:proposedapproach}, we present the proposed approach, followed by the evaluation and analysis of the results of our experiments in Section~\ref{section:evaluation}. Finally, Section~\ref{section:conclusion} summarizes the key findings, discussing their implications and suggesting potential directions for future work. 
\begin{table*}[t]
\caption{Summary of Related Works on Migration Strategies in Cloud-Assisted Edge Environments}
\label{table:related-works}
\resizebox{\textwidth}{!}{
\begin{tabular}{|c|c|c|c|c|c|c|c|c|} \hline  

                       & \multicolumn{2}{|c|}{Environment}                                 & \multicolumn{2}{|c|}{Factors}                                                             & \multicolumn{4}{|c|}{Migration Strategy}\\ \hline  
Paper                  & \multicolumn{1}{|c|}{Kubernetes} & Cloud-Assisted Edge Cluster & \multicolumn{1}{|c|}{Fragmentation} & \multicolumn{1}{|c|}{Quality of Service} & Container Placement & \multicolumn{1}{|c|}{Cloud to Edge} & Edge to Cloud  & Edge to Edge\\ \hline  
\text{\cite{li_heterogeneity-aware_2020}}       & \xmark           & \cmark& \xmark              & \cmark                   & \xmark& \xmark              & \cmark&\xmark\\ \hline  
\text{\cite{ma_cost-efficient_2021}}       & \xmark           & \cmark& \xmark              & \cmark                   & \xmark& \xmark              & \cmark&\xmark\\ \hline  
\text{\cite{wang_dynamic_2017}}     & \xmark           & \cmark& \xmark              & \cmark                   & \xmark& \cmark              & \xmark&\cmark\\ \hline  
\text{\cite{kim_optimal_2021}}      & \xmark           & \xmark& \xmark              & \cmark                   & \xmark& \xmark              & \xmark&\cmark\\ \hline  
\text{\cite{chen_service_2023}}     & \xmark           & \xmark& \xmark              & \cmark                   & \cmark& \xmark              & \xmark&\cmark\\ \hline  
\text{\cite{li_re-scheduling_2023}}       & \xmark           & \xmark& \xmark              & \cmark                   & \cmark& \xmark              & \xmark&\cmark\\ \hline  
\text{\cite{chi_multi-criteria_2023}}      & \cmark           & \cmark& \xmark              & \xmark                   & \xmark& \cmark              & \xmark&\cmark\\ \hline  
\text{\cite{rong_live_2023}}     & \cmark           & \xmark& \xmark              & \xmark                   & \xmark& \xmark              & \xmark&\cmark\\ \hline  
\text{\cite{ghafouri_mobile-kube_2022}} & \cmark           & \xmark& \xmark              & \xmark                   & \cmark& \xmark              & \xmark&\cmark\\ \hline  
\text{\cite{lai_delay-aware_2023}}      & \cmark           & \xmark& \xmark              & \xmark                   & \cmark& \xmark              & \xmark&\cmark\\ \hline  
\text{\cite{marchese_2022}}      & \cmark           & \cmark& \cmark              & \cmark                   & \xmark& \xmark              & \xmark&\xmark\\ \hline  
\text{\cite{chiaro_2024}}      & \cmark           & \cmark& \xmark              & \cmark                   & \cmark& \cmark              & \cmark&\cmark\\ \hline  
\text{\cite{rausch_2021}}      & \cmark           & \cmark& \xmark              & \cmark                   & \xmark& \xmark              & \xmark&\xmark\\ \hline  
\text{\cite{qiao_2024}}      & \cmark           & \cmark& \xmark              & \cmark                   & \xmark& \xmark              & \xmark&\xmark\\ \hline  
\text{\cite{ding_2023}}      & \cmark           & \xmark& \cmark              & \cmark                   & \xmark& \xmark              & \xmark&\xmark\\ \hline  
Proposed Approach      & \cmark  & \cmark& \cmark     & \cmark          & \cmark& \cmark     & \cmark&\cmark\\ \hline 
\end{tabular} 
}
\end{table*}

\section{Related Work}
\label{section:relatedwork}
The field of edge computing has evolved significantly in recent years, driven by the need for efficient resource management and low-latency services. In this section, we focus on two key research areas: the first subsection discusses strategies and techniques for resource management in edge clusters, while the second subsection examines efforts to tailor Kubernetes, the leading container orchestration platform, to the unique requirements of edge computing environments. \par
\subsection{Resource Management in Edge Clusters}
To optimize resource usage on edge clusters and hence the QoS, several scheduling and migration techniques have been proposed in the literature. The authors in \cite{ma_cost-efficient_2021} addressed scalability in edge platforms with the Cloud Assisted Mobile Edge (CAME) framework by outsourcing mobile requests to cloud instances, which accommodates dynamic requests and various quality of service requirements. They proposed Optimal Resource Provisioning (ORP) algorithms to optimize edge computation capacity and dynamically adjust cloud tenancy. Evaluations showed these algorithms outperform local-first and cloud-first benchmarks in flexibility and cost-efficiency. Li et al. \cite{li_heterogeneity-aware_2020} studied a cloud-assisted edge computing system (CAECS) to address the challenges of edge computing, such as handling randomly varying workloads. They proposed a replica placement strategy to meet diverse user demands and reduce response time, and a data migration strategy to ensure data reliability. Additionally, a heterogeneity-aware elastic provisioning strategy was introduced to manage cloud instance rentals. The authors in \cite{kim_optimal_2021} introduced three migration algorithms and developed an algorithm-selector mechanism that chooses the most appropriate algorithm based on the characteristics of the container. This approach enabled them to achieve live migration durations of less than one second in their tests. Moreover, \cite{chen_service_2023, li_re-scheduling_2023} approached migration by modeling it as partially observable Markov decision processes and employing multi-objective, multi-constraint optimization techniques. This enabled them to achieve improvements in average delay and reductions in power consumption. Similarly, in \cite{wang_dynamic_2017}, Wang et al. proposed an actual cost predictive model. By assuming pre-determined upper bounds, they were able to make sub-optimal placement decisions based on their predicted costs in their simulations. \par
Additionally, several studies have leveraged service migration to enhance service quality in mobile edge clusters. For example, Chi et al. \cite{chi_multi-criteria_2023} proposed a method for live migration of services in mobile edge clusters using a multi-criteria decision-making algorithm based on TOPSIS. This method optimizes migration processes and alleviates traffic congestion. Furthermore, Rong et al. \cite{rong_live_2023} introduced three approaches for migrating video analysis applications within edge clusters, ensuring minimal noticeable impact on service quality for users during the transition.  \par
Moreover, several recent studies have explored resource management techniques specifically targeting cloud-assisted edge clusters. Marchese et al. \cite{marchese_2022} proposed a network-aware scheduler and descheduler for Kubernetes environments, optimizing pod placement and migration based on real-time network conditions and microservice communication patterns. Their approach effectively reduces communication latency within distributed edge-cloud infrastructures. Also, Chiaro et al. \cite{chiaro_2024} introduced LAIS, a latency-aware scheduling framework that dynamically schedules and migrates pods across multi-cluster edge-cloud environments based on user-defined latency constraints and real-time measurements. Their method significantly improves QoS while adapting to changing network conditions. \par
\subsection{Adapting Kubernetes to Edge clusters}
Numerous works address the challenges related to Kubernetes adaptation within edge infrastructure. For instance, Ghafouri et al. \cite{ghafouri_mobile-kube_2022} proposed a reinforcement learning-based scheduler aimed at reducing energy consumption. This study demonstrated that learning-based solutions could effectively replace traditional algorithmic approaches through innovative methods. Similarly, Lai et al. \cite{lai_delay-aware_2023} introduced a solution that combines a scheduler and a scorer. The scheduler uses the scorer in a filtering stage to select nodes with minimum network latency and the most available resources to serve the pods, thereby achieving a balance between processing delays and network latency among pods. Moreover, Zhang et al. \cite{zhang_effective_2023} analyzed the user's probability function of sojourn time to characterize user mobility intensity and service deployment overhead. The resulting scheduler was able to reduce user-perceived latency, constrain service migration, and optimize user experience quality in software-defined (SDN) networks. \par
Recent works have further advanced Kubernetes scheduling and orchestration mechanisms in cloud-assisted edge environments. Rausch et al. \cite{rausch_2021} proposed Skippy, a data locality-aware scheduler designed to improve serverless function placement in heterogeneous edge-cloud infrastructures by considering data proximity, resource capabilities, and network bandwidth. Qiao et al. \cite{qiao_2024} developed EdgeOptimizer, a modular Kubernetes-based scheduling framework that enables programmable orchestration of time-critical tasks while supporting customizable scheduling algorithms for edge-cloud scenarios. Chiaro et al. Furthermore, Ding et al. \cite{ding_2023} presented a Kubernetes-oriented placement model using genetic algorithms for dynamic resource allocation and availability-aware microservice deployment, aiming to minimize resource and communication costs in dynamic edge environments. \par
Efficient utilization of edge resources reduces the necessity of offloading computations from edge servers to the cloud during resource shortages, thus lowering the average latency of applications running on the edge platform with cloud cluster assistance. Although several studies, like those mentioned earlier, have addressed migration and resource allocation in edge clusters, they have not focused on reducing latency by optimizing the utilization of edge nodes through simultaneous service migration and dynamic scheduling in Kubernetes-managed clusters. To the best of our knowledge, this is the first work that considers both migration and scheduling simultaneously on cloud-assisted edge clusters. Our approach aims to achieve efficient resource allocation in Kubernetes-managed environments while accommodating the cluster's dynamic nature through pod migration. Table~\ref{table:related-works} summarizes the related works. \par

\section{System model}
\label{section:systemmodel}
\begin{figure*}[!ht]
    \centering
    \includegraphics[width=0.8\linewidth]{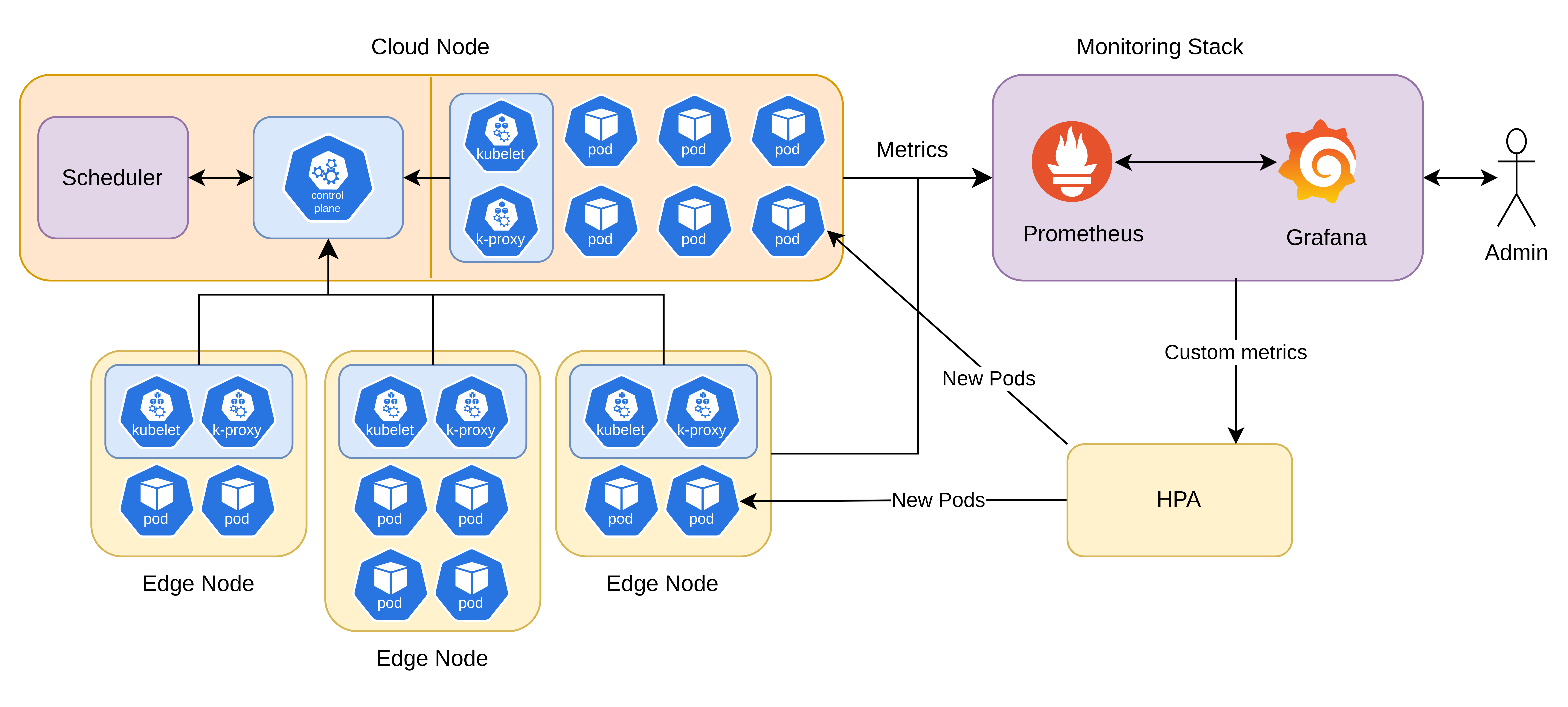}
    \caption{Kubernetes-managed cloud-assisted edge cluster model}
    \label{fig:cluster-model}
\end{figure*}

Edge clusters are typically composed of resource-constrained, heterogeneous, failure-prone nodes placed at the network's edge. These nodes enable the processing of user requests with reduced latency. To mitigate these limitations, edge clusters are typically paired with cloud clusters that feature nodes with greater resources and fault-tolerant architectures. This forms an extensive system of cloud and edge nodes, commonly referred to as cloud-assisted edge clusters. This configuration enhances the overall capability and reliability of the network, ensuring efficient processing and improved service delivery. Figure~\ref{fig:cluster-model} illustrates the cloud-assisted edge cluster with its components. \par

Kubernetes, the de facto standard for container orchestration, efficiently manages application deployments and their life cycles within a cluster. As depicted in Figure~\ref{fig:cluster-model}, several essential components are crucial for Kubernetes to perform this task. These components typically include a scheduler for resource allocation, KubeProxy for network management, a monitoring system for performance tracking, and horizontal pod autoscaling (HPA) for dynamic scaling of application instances. \par
\subsection{Scheduler}
The scheduler is a vital component of the Kubernetes control plane, responsible for assigning newly created pods to nodes. It operates through a filtering and scoring phase to determine the optimal node for pod binding. The default scheduler, known as kube-scheduler, is installed alongside other control plane components in Kubernetes \cite{noauthor_kube-scheudler_nodate}. By default, the kube-scheduler aims to evenly distribute pods across nodes to enhance availability and reliability \cite{noauthor_kube-scheduler-eviction_nodate}.

\subsection{KubeProxy}
KubeProxy is a key Kubernetes component that runs on each node and manages network rules to ensure connectivity between services and pods. This service allows for seamless communication and access to pods both from within and outside the cluster, facilitating efficient interaction across the entire network.

\subsection{Kubelet}
Kubelet is another crucial component of Kubernetes that runs on each node within the cluster. It ensures that containers are running in a pod by interacting with the container runtime and the API server. Kubelet receives PodSpecs from the API server and ensures that the containers described are running and healthy. By constantly monitoring the state of pods and containers, Kubelet plays a key role in maintaining the desired state of the cluster and ensuring that applications are running as intended.

\subsection{Monitoring Stack}
Given the failure-prone nature of edge nodes, a robust monitoring stack is crucial for continuously tracking each node's state and integrating this data into various system behaviors. Prometheus \cite{noauthor_prometheus_nodate} is widely used for metric recording, often deployed in conjunction with Grafana \cite{noauthor_grafana_nodate} for metric visualization and observability. This combination provides a comprehensive monitoring solution that is commonly adopted in the industry to ensure the health and performance of edge infrastructure.

\subsection{Horizontal Pod Autoscaler (HPA)}
Horizontal Pod Autoscaler (HPA) is a crucial API resource in Kubernetes that dynamically adjusts the number of pods for a service based on observed CPU utilization, memory usage, or custom metrics obtained from the monitoring stack. This functionality enables the Kubernetes cluster to respond to fluctuating workloads and optimize resource allocation by altering the number of replicas as needed. Consequently, HPA ensures efficient resource utilization and maintains application performance within the Kubernetes environment.
\begin{table}[t]
\caption{Summary of all notations}
\label{table:notations}
\begin{tabularx}{\columnwidth}{|c|X|}
\hline
Symbol                 & Definition                                                                                                      \\ \hline
$E$                      & Set of all edge nodes in the edge cluster                               \\ \hline
$E_i$                      & $i$th node in the edge cluster                                                              \\ \hline
$R_{E_i}$            & Resource vector of edge node $E_i$                                                                                      \\ \hline
$t_C$                & Request latency when executed on cloud                                                                         \\ \hline
$t_E$                & Request latency when executed on edge nodes                                                                          \\ \hline
$S$                      & Set of all deployments                           \\ \hline
$P_C$              & The set of all cloud pods                 \\ \hline
$P_E$              & The set of all edge pods                                                                       
\\ \hline
$P_{i,j}$              & The $j$th pod of the $i$th service                                                                          \\ \hline
$R_{S_i}$            & Resource vector for pods of service $S_i$                                                                         \\ \hline
$Q_{i}$              & Quality of service guaranteed for service $S_i$                                                                    \\ \hline
$N$                      & Set of all newly created pods, which is input to the scheduler                                                        \\ \hline
\(u\)                      & Allocation mapping for all pods                                                                                  \\ \hline
\(u'\)                 & Allocation mapping for all pods before scheduler decision                                                        \\ \hline
\(u_{i,j,k}\)          & Whether the pod $P_{i,j}$ scheduled on the $k$th node \\ \hline
$M_{C2E}$          & The maximum number of pods that the scheduler will migrate from the cloud to the edge in each suggestion \\ \hline
$M_{ER}$          & The maximum number of pods that the scheduler will reorder in the edge (migrate from one edge node to another edge node) in each suggestion or scheduling request \\ \hline
$M_{CPU}$          & The maximum amount of CPU cores provided by the edge nodes \\ \hline
$M_{MEM}$          & The maximum amount of memory in GB provided by the edge nodes \\ \hline 
\end{tabularx}
\end{table}

\section{Problem Definition}
\label{section:problemdefinition}
This section defines the scheduling and migration problems that the scheduler aims to solve. Before defining the problems, it is necessary to state the assumptions the scheduler considers from its environment, input, and output. 
\subsection{Assumptions}
To reduce the complexity of the scheduling space, we have made the following assumptions.
\subsubsection{Cluster Assumptions}
We assume that there is a single Kubernetes cluster, consisting of both edge nodes and cloud nodes. The Kubernetes scheduler is aware of the node types. The edge and cloud nodes are represented by the sets $E$ and $C$ respectively. Without loss of generality, we assume a single cloud node with infinite resource capacity in this work.

The edge nodes are heterogeneous, and the resources of the $i$th node are represented as a two-dimensional vector $R_{E_i}$, comprising the number of processor cores, and the volume of memory of this node in gigabytes. 
\subsubsection{Service Assumptions}
The scheduler assumes that users may request one of the services from the service set $S$, each comprising several pods that aim to respond to user requests. The set $P_i$ denotes the pods of the $i$th service deployment $S_i$. The scheduler is not responsible for determining the number of pods for each service, but must make placement decisions based on the assumption that this number can vary by HPA. It is also assumed that all pods of service $S_i$ require the same resources presented as a two-dimensional vector $R_{S_i}$, where the dimensions correspond to the number of processor cores, and the volume of memory required by that service. \\
Furthermore, the service provider provides a QoS level for each service, stating that at least $Q_i$ fraction of $P_i$ should be deployed on edge.

\subsubsection{Scheduler Assumptions}
At any given moment, the scheduler has a collection, potentially with duplicate members, denoted by $N$, representing the deployments of newly created pods. The scheduler outputs a binary allocation matrix, $u$, where $u_{i,j,k}=1$ if the pod $P_{i,j}$ has been placed on the node $E_k$ in the edge cluster. \\
All the notations are summarized in Table \ref{table:notations}.

\subsection{Problem Statement}
The main objective of the scheduler is to keep users satisfied based on the given guarantees on QoS, i.e. maximizing $QoS(u)$. As the secondary objective, the scheduler tries to perform the minimum number of migrations. We define the overall quality of service ($QoS(u)$) as the accumulated QoS of all services as follows:
\begin{equation}
\label{eq:qos-overall}
    QoS(u)=\sum_{i \leq|S|}{QoS(i,u)}
\end{equation}
where $QoS(i,u)$ is the QoS level of service $S_i$. The scheduler aims to establish the maximum number of QoS guarantees possible; if it is not possible, it strives to get as close as possible. To do this, we define $\Delta(i,u)$ as the difference between the current level of QoS and the expected level of QoS ($Q_i$) for service $S_i$.
\begin{equation}
\label{eq:delta}
    \Delta(i,u)=\frac{\sum_{j\leq|P_i|,k\leq|E|}{u_{i,j,k}}}{\left|P_i\right|}-Q_i
\end{equation}

Then, we define $F$ and $QoS(i,u)$ as follows: 
\begin{equation}
\label{eq:qos-F}
    F(x) = 
    \begin{cases}
      \alpha x & \text{if } x < 0 \\
      \beta x + \gamma & \text{if } x \geq 0
    \end{cases}
\end{equation}
\begin{equation}
\label{eq:qos-service}
    QoS(i,u)=F(\Delta(i))
\end{equation}
The $F$ function is designed to transform the difference $\Delta(i,u)$ into a QoS value that reflects the satisfaction level of each service. When $\Delta(i,u)$ is negative, the current QoS is below the expected level, hence we apply a linear penalty with severity of $\alpha$. This motivates the scheduler to increase the QoS level, if it does not meet the expected level. When it is non-negative, meaning the current QoS meets or exceeds the expected level, we increase the reward by a large value $\gamma$ to incentivize the scheduler to maximize QoS. We also apply a linear reward by the rate of $\beta$ to motivate the scheduler to keep increasing the QoS level, even if all QoS requirements are met. Overall, these constants follow the inequality below:
\begin{equation}
\label{eq:const-ineq}
    0 \leq \beta < \alpha << \gamma
\end{equation}
After each scheduling event, the scheduler changes the allocation mapping from $u'$ to $u$. The number of migrations performed for the pod $P_{i, j}$ that is not in $N$ can be defined as follows:
\begin{equation}
\label{eq:migrations}
    \theta(i,j)=\frac{
        \sum_{k\leq |E|}{\left|u_{i,j,k}-u'_{i,j,k}\right|}+\left|\sum_{k\leq |E|}{u_{i,j,k}}-\sum_{k\leq |E|}{u'_{i,j,k}}\right|
    }{
    2
    }
\end{equation}
The first term of the equation \ref{eq:migrations} counts the number of changes in the edge allocation of pods (the value will be increased by two for each migration between edge nodes and by one for edge-to-cloud/cloud-to-edge migration). The second term calculates how many pods have migrated from cloud to edge and vice versa. As each migration is counted twice, the overall value is divided by two. \\

The following are the problem constraints:
\begin{enumerate}
    \item Pod-to-node allocation constraints: Each pod can be assigned to at most one of the cluster nodes. Hence, we have:
    \begin{equation}
    \label{eq:alloc-constr}
    \begin{aligned}
        & u_{i,j,k}\in \{0,1\},\\
        & \sum_{k\leq |E|}{u_{i,j,k}} \leq 1 \qquad \forall i\leq|S|,j\leq |P_i|
    \end{aligned}
    \end{equation}
    When a pod is not allocated to any edge node (i.e. $\sum_{k\leq |E|}{u_{i,j,k}}=0$), the pod is deployed on the cloud.
    
    \item Resource constraints: The sum of resources required by the pods allocated to each node must be less than that node's resources:
    \begin{equation}
    \label{eq:resource-constr}
           \sum_{i\leq |S|,j\leq |P_i|}{u_{i,j,k}\times R_{s_i}} \leq R_{E_k}
           \qquad
           \forall k\leq |E|
    \end{equation}
\end{enumerate}
The scheduler aims to maximize user QoS (primary goal) and minimize migrations count (secondary goal) subject to the above constraints:
\begin{equation}
\begin{aligned}
    & \max_{u \in U} \sum_{i \leq |S|}{QoS(i,u)}  \\
    & \min_{u \in U} \sum_{i \leq |S|,j \leq |P_i|, P_{i, j} \notin N}{\theta(i,j)}
\end{aligned}
\end{equation}
The presented problem is a MILP\footnote{Mixed-integer linear programming} optimization problem over integers, and its optimal solution is NP-hard. We approach solving this problem by proposing KubeDSM, a Kubernetes scheduler that incorporates various combinatorial methods.
\begin{table*}[t]
\centering
\caption{Possible steps, actions and their verification}
\label{tab:possible-actions}
\resizebox{\textwidth}{!}{
    \begin{tabular}{|l|l|l|}
    \hline
    \multicolumn{1}{|l|}{Step}    & \multicolumn{1}{l|}{Action}                     & \multicolumn{1}{l|}{Verification}            \\ \hline
    Create a pod for a deployment & Nothing (will be created by HPA)                & A pod creation event for the same deployment \\ \hline
    Bind a pod to a node          & Submit a pod-target binding request to the API server & A pod changed event for the desired pod with the pod’s node being the target node \\ \hline
    Delete a pod                  & Submit a pod deletion request to the API server & A pod deletion event for the desired pod     \\ \hline
    \end{tabular}
}
\end{table*}

\section{Proposed Approach}
\label{section:proposedapproach}
In this section, we present in detail the system architecture of the proposed KubeDSM scheduler, including its key components, their functionalities, and the proposed scheduling algorithms.
\subsection{Scheduler Overview}
\begin{figure}[!ht]
    \centering
    \includegraphics[width=1.0\linewidth]{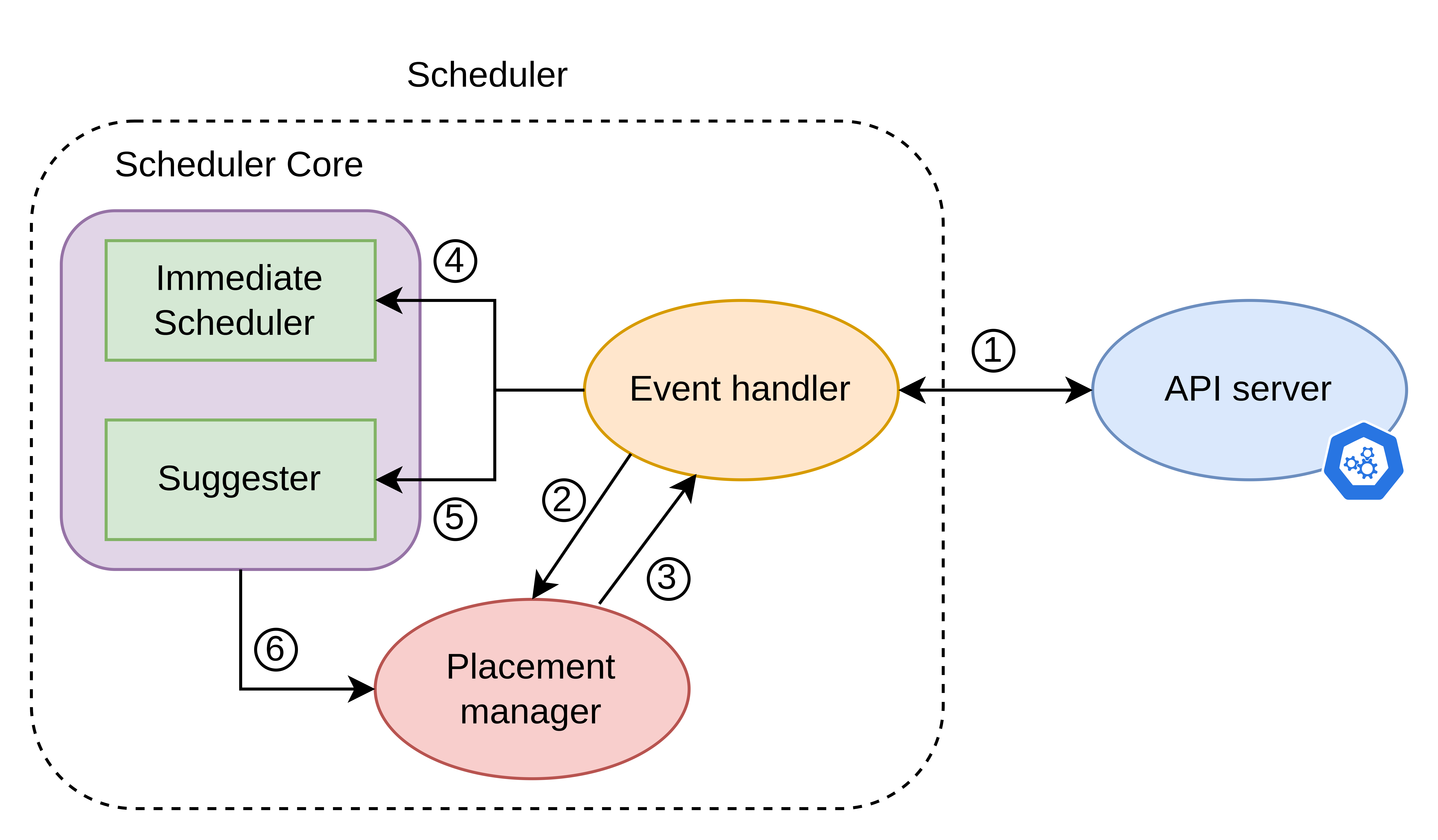}
    \caption{KubeDSM component diagram}
    \label{fig:component-diagram}
\end{figure}
As illustrated in Figure \ref{fig:component-diagram}, the scheduler is composed of three main components.

\textbf{Event Handler:} The event handler serves as the communication bridge between the scheduler and Kubernetes. It is subscribed to the events channel on the Kubernetes API server to capture all events that occur within the scheduler’s namespace \circled{1}. In addition, it periodically requests information about all nodes and pods to account for missed or unexpected events. Each event received is forwarded to the Placement Manager component \circled{2}, which then sends a response detailing the action to be taken \circled{3}. Possible actions include waiting for another event, ignoring the current event, or performing tasks such as deleting a pod or deploying a pod on a specific node. Using the information gathered from events or periodic requests, the event handler constructs and maintains a cluster state (CS), representing the scheduler's understanding of the current cluster state. The event handler continuously updates this cluster state over time.

\textbf{Placement Manager:} This component is responsible for implementing current pod placement plans. Each plan comprises a sequence of steps, and if any step fails or is canceled, the subsequent steps will also be canceled, resulting in only partial execution of the plan. A step includes two parts:
\\
\textit{Action}, which is the specific action required to execute the step, and
\\
\textit{Verification}, that is the method used to verify that the action was successfully and completely executed. Verification is achieved by receiving a specific type of event from the event handler.

Table \ref{tab:possible-actions} outlines the possible steps.
When a new event is received from the event handler, one of the following scenarios occurs:
\begin{itemize}
    \item The event matches the current state of one of the plans: the step is verified, and the next step in that plan is executed (or the plan is completed).
    \item The event is incompatible with some plans (e.g., it pertains to the same pod, but the type or expected information does not match): the related plans are canceled. In this case, any remaining pending pods will be deployed to the cloud.
    \item The event concerns the creation of a new pod: it is forwarded to the scheduler core (first via step \circled{3}, then step \circled{4}).
    \item The event is deemed irrelevant, so it is safely ignored.
\end{itemize}

\textbf{Scheduler Core:} The scheduler core is composed of two main components:
\\
\textit{Immediate scheduler}, this component is responsible for scheduling newly created pods. It receives a list of new pods from the event handler \circled{4} and generates plans based on the scheduling algorithm to bind the pods to appropriate nodes. This process does not involve migration, resulting in single-step plans that only focus on binding new pods to selected nodes.
\\
\textit{Suggester}, this component is responsible for suggesting plans for pod reordering. It is periodically called by the event handler \circled{5} to provide migration suggestions. These suggestions are formed by creating multiple plans based on suggestion algorithms for pod migration, aiming to place more pods on edge nodes and optimize the utilization of edge resources.
\\
The plans generated by both components are then forwarded to the Placement Manager \circled{6}.
\\
The scheduler defines the following constants (as its configuration) and utilizes them in the algorithm:
\begin{itemize}
    \item $M_{C2E}$: The maximum number of pods that the scheduler will migrate from the cloud to the edge in each suggestion.

    \item $M_{ER}$: The maximum number of pods that the scheduler will reorder in the edge (migrate from one edge node to another edge node) in each suggestion or scheduling request.

    \item $M_{CPU}$ and $M_{MEM}$: The maximum amount of resources provided by the edge nodes (CPU cores and memory in gigabytes respectively).
\end{itemize}

\subsection{Algorithms overview}
\begin{figure}[!h]
    \centering
    \includegraphics[width=1.0\linewidth]{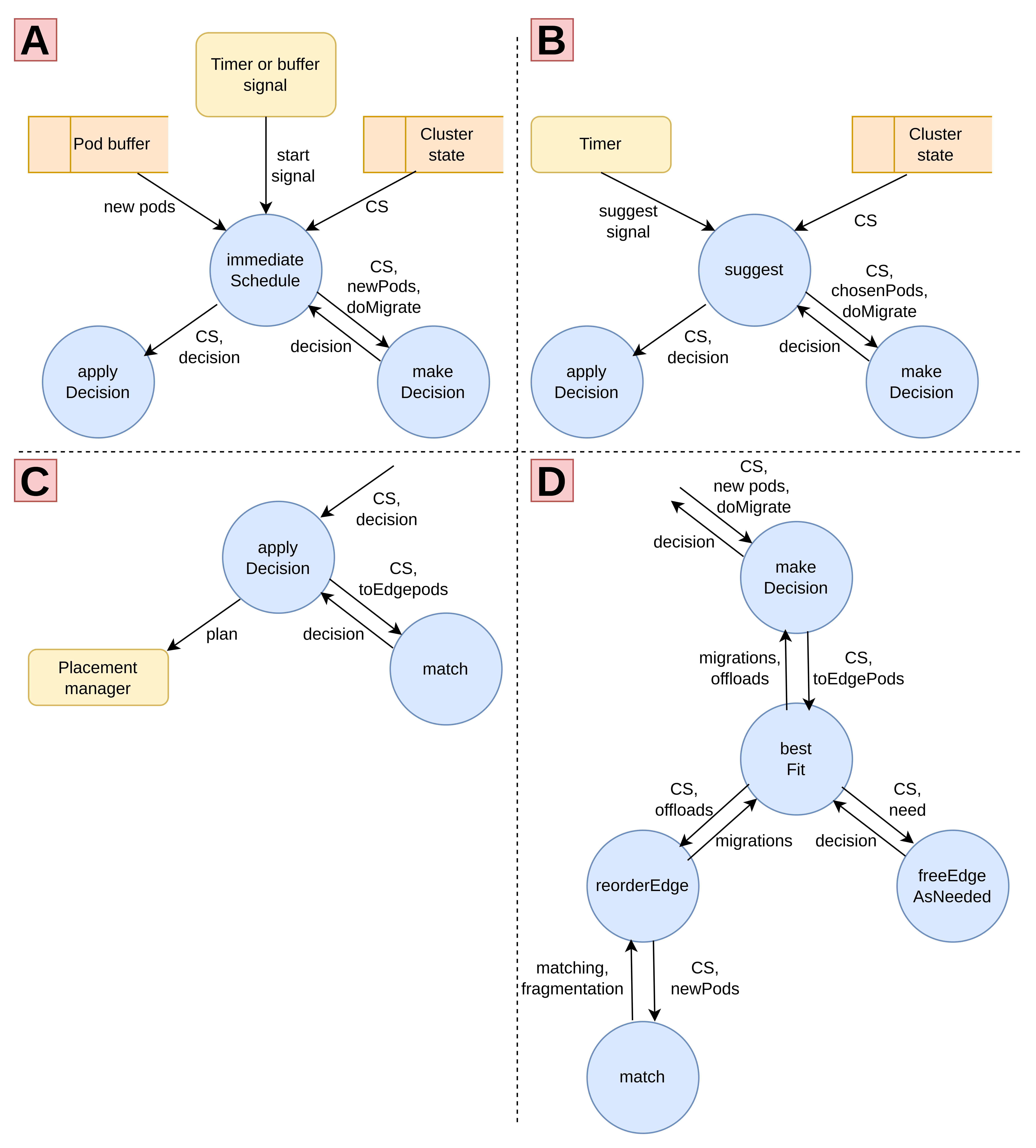}
    \caption{Scheduler Complete Call Graph}
    \label{fig:algorithm-DFD}
\end{figure}
Figure \ref{fig:algorithm-DFD} demonstrates the data flow diagram of the scheduler’s algorithm. It features two entry points: \verb|suggest| and \verb|ImmediateSchedule|. The remaining parts of the algorithm, presented in reverse dependency order, include \verb|match|, \verb|reorderEdge|, \verb|freeEdgeAsNeed|, \verb|bestFit|, \verb|makeDecision|, and \verb|applyDecision|. We describe each component in this order.

\textbf{immediateSchedule:} As shown in Algorithm \ref{alg:immediate_schedule}, this algorithm is a simple call to \verb|makeDecision| with the list of new pods, with migration disabled (as new pods need to be scheduled as quickly as possible to avoid potential connection loss for certain applications). Following this, it invokes \verb|applyDecision| with the generated decision.
\begin{algorithm}
\caption{Immediate Schedule Algorithm}\label{alg:immediate_schedule}
\KwIn{Cluster State (CS), newPods}
\SetKwFunction{FImmediateSchedule}{ImmediateSchedule}
\SetKwProg{Fn}{Function}{:}{}
\Fn{\FImmediateSchedule{CS, newPods}}{
    decision $\gets$ makeDecision(CS, newPods, \textbf{false})\;
    applyDecision(CS, 
 decision)\;
}
\end{algorithm}

\textbf{suggest:} The goal of the suggest algorithm (Algorithm \ref{alg:suggest}) is to suggest some migrations to offload (migrate) as many pods as possible from the cloud to the edge (to increase $QoS$) and reorder (migrate) some pods within the edge cluster, to decrease fragmentation if needed. It begins by creating a list of pods to be placed on the edge nodes. It then makes a decision using the selected pods as new pods, with migration enabled. Finally, it applies the decision. As shown in Algorithm \ref{alg:suggest}, for creating the list, in each iteration, as long as the number of selected pods is less than $M_{C2E}$ and there is at least one candidate pod in the cloud, it sorts the candidate pods by their score and checks the pod with the highest score. If the pod, along with the already selected pods, fits on the edge, it is added to the list. The pod is then removed from the candidate list, and the iterations continue.
The scoring function is defined as:
\begin{equation}
\label{eq:suggest-score}
    \begin{aligned}
        &score(pod)= \\ 
        &\frac{QoS(CS'-\{pod\ on\ cloud\} \cup \{pod\ on\ edge\})-QoS(CS')}{size(pod)}
    \end{aligned}
\end{equation}
The score function aims to capture the increase in $QoS$ per unit of resource moved to the edge. The increase in $QoS$ is calculated as the difference between when the pod is on the edge and on the cloud. To find the increase per resource unit moved, score is divided by the pod's size, which is calculated as:
\begin{equation}
    size(pod)=\sqrt{
        \frac{pod_{CPU}}{M_{CPU}}
        \times
        \frac{pod_{MEM}}{M_{MEM}}
    }
\end{equation}

\begin{algorithm}
\caption{Suggest Algorithm}\label{alg:suggest}
\KwIn{Cluster State (CS)}

\SetKwFunction{Fsuggest}{suggest}
\SetKwProg{Fn}{Function}{:}{}
\Fn{\Fsuggest{CS}}{
    candidPods $\gets$ all cloud pods\;
    chosenPods $\gets$ []\;
    CS' $\gets$ CS \tcp*[r]{a copy of cluster state}
    currentFreeResources $\gets$ freeResources in edge\;
    \While{len(chosenPods) $<$ $M_{C2E}$ \textbf{and} len(candidPods) $>$ 0}{
        sort candidPods by score (Eq \ref{eq:suggest-score}) in decreasing order\;
        firstCandidate $\gets$ candidPods.pop()\;
        \If{firstCandidate.resources $\leq$ currentFreeResources}{
            chosenPods += firstCandidate\;
            currentFreeResources -= firstCandidate.resources\;
            CS' $\gets$ CS' - \{firstCandidate on cloud\} $\cup$ \{firstCandidate on edge\}\;
        }
    }
    decision $\gets$ makeDecision(CS, chosenPods, \textbf{true})\;
    applyDecision(CS, decision)\;
}
\end{algorithm}

\textbf{makeDecision:} As demonstrated in algorithm \ref{alg:make_decision}, the \verb|makeDecision| function begins by choosing to deploy all new pods to the cloud as the default decision. It then iteratively evaluates each subset of new pods, denoted as $toEdge$, to determine the possibility of deploying $toEdge$ on edge nodes. If the combined resources of $toEdge$ exceed the total available edge resources, that subset is discarded. If migration is allowed, the function calculates the optimal and minimal set of migrations required to accommodate $toEdge$ on edge using the \verb|bestFit| function. If migration is not allowed, feasibility is checked approximately by comparing the total resources of $toEdge$ against the available resources in the edge nodes. Finally, the Quality of Service (QoS) score for both the current decision and the best decision is evaluated. If the current decision yields a higher QoS score than the previously recorded best decision, it is selected as the new best decision. The function ultimately returns the best decision found through this process.

\begin{algorithm}
\caption{Make Decision Algorithm}\label{alg:make_decision}
\KwIn{Cluster State, newPods, doMigrate}
\KwOut{bestDecision}

\SetKwFunction{FmakeDecision}{makeDecision}
\SetKwProg{Fn}{Function}{:}{}
\Fn{\FmakeDecision{CS, newPods, doMigrate}}{
    bestDecision $\gets$ place all new pods on cloud\;
    \ForEach{subset $toEdge \subset newPods$}{
        \If{sum of resources of $toEdge$ $>$ edge's resources}{
            \textbf{continue}\;
        }
        currentDecision $\gets$ deploying $toEdge$ on edge and other new pods on cloud\;
        \If{doMigrate}{
            migrationsNeeded $\gets$ bestFit(CS, $toEdge$)\;
            currentDecision += performing migrationsNeeded\;
        }
        \Else{
            \If{sum of resources of $toEdge$ $>$ free resources in edge}{
                \textbf{continue}\;
            }
        }
        \If{QoS(CS after bestDecision) $>$ QoS(CS after currentDecision)}{
            bestDecision $\gets$ currentDecision\;
        }
    }
    \Return{bestDecision}\;
}
\end{algorithm}

\textbf{bestFit:} Algorithm \ref{alg:best_fit} aims to determine the optimal set of migrations needed to free up a specified amount of resources on the edge (referred to as \textit{need}). The term "optimal" here means: First, to meet the required amount of freed resources on the edge, the algorithm prioritizes minimizing the associated QoS loss. Second, among the possible sets of offloads that achieve the same QoS, the algorithm selects the set that involves the fewest number of migrations. After identifying the best set of migrations to free up edge resources, the algorithm then attempts, if possible, to perform up to $M_{ER}$ migrations. The goal is to reorder the edge in a manner that reduces fragmentation, making it easier for future pods to be accommodated on the edge nodes.

\begin{algorithm}
\caption{Best Fit Algorithm}\label{alg:best_fit}
\KwIn{Cluster State (CS), toEdgePods}
\KwOut{migrations}

\SetKwFunction{FbestFit}{bestFit}
\SetKwProg{Fn}{Function}{:}{}
\Fn{\FbestFit{CS, toEdgePods}}{
    offloads $\gets$ freeEdgeAsNeeded(CS, sum resources in toEdgePods) \tcp*[r]{Offloads are migrations from the edge to the cloud to increase $QoS$.}
    CS' $\gets$ CS after offloads\;
    reorderings $\gets$ reorderEdge(CS') \tcp*[r]{Reorderings are migrations inside the edge for fragmentation reduction.}

    \Return{offloads $\cup$ reorderings}\;
}
\end{algorithm}

\textbf{freeEdgeAsNeeded:} The \verb|freeEdgeAsNeeded| algorithm (Algorithm \ref{alg:free_edge_as_needed}) operates similarly to the \verb|suggest| algorithm, but in reverse. Its goal is to minimize score loss while freeing up the required amount of resources on the edge. It sorts the edge pods using a similar scoring function as the \verb|suggest| algorithm and proceeds to free them one by one until the needed resources are released. Since the scoring function changes each time a pod is freed, it is crucial to re-sort the candidate list after each pod is removed.
The score function is defined as follows:
\begin{equation}
\label{eq:free-edge-as-needed-score}
    \begin{aligned}
        &score(pod)= \\ 
        &\frac{QoS(CS'-\{pod\ on\ edge\} \cup \{pod\ on\ cload\})-QoS(CS')}{size(pod)}
    \end{aligned}
\end{equation}

\begin{algorithm}
\caption{Free Edge As Needed Algorithm}\label{alg:free_edge_as_needed}
\KwIn{Cluster State, need}
\KwOut{freedPods}

\SetKwFunction{FfreeEdgeAsNeeded}{freeEdgeAsNeeded}
\SetKwProg{Fn}{Function}{:}{}
\Fn{\FfreeEdgeAsNeeded{CS, need}}{
    candidPods $\gets$ all edge pods\;
    freedPods $\gets$ []\;
    CS' $\gets$ CS \tcp*[r]{a copy of cluster state}
    currentFreeResources $\gets$ freeResources in edge\;
    \While{len(candidPods) $>$ 0 \textbf{and} currentFreeResources $<$ need}{
        sort candidPods by score in decreasing order\;
        firstCandidate $\gets$ candidPods.pop()\;
        freedPods += firstCandidate\;
        currentFreeResources += firstCandidate.resources\;
        CS' $\gets$ CS' - \{firstCandidate on edge\} $\cup$ \{firstCandidate on cloud\}\;
    }
    \Return{freedPods}\;
}
\end{algorithm}

\textbf{reorderEdge:} This function (Algorithm \ref{alg:reorder_edge}) aims to reduce fragmentation within the edge nodes by identifying potential pod migrations within the edge. The resource fragmentation for a cluster stage is calculated based on the following equation:
\begin{equation}
\label{eq:fragmentation}
    \begin{aligned}
        &frag(CS)= \\ 
        &\sum_{\text{all nodes}} \left( 1 - \frac{\text{$CPU_{Used}$}}{\text{$CPU_{Total}$}} \times \frac{\text{$MEM_{Used}$}}{\text{$MEM_{Total}$}} \right)
    \end{aligned}
\end{equation}
The algorithm evaluates each subset of edge pods, up to a size of $M_{ER}$, and considers migrating that subset. Using the \verb|match| algorithm, it determines the optimal target nodes for each pod. The resulting fragmentation is then calculated for each subset and the subset with the least fragmentation, along with its best target nodes, is selected as the optimal set of reorder migrations.

\begin{algorithm}
\caption{Reorder Edge Algorithm}\label{alg:reorder_edge}
\KwIn{Cluster State}
\KwOut{bestMigrations}

\SetKwFunction{FreorderEdge}{reorderEdge}
\SetKwProg{Fn}{Function}{:}{}
\Fn{\FreorderEdge{CS}}{
    bestMigrations $\gets$ \{\}\;
    leastFrag $\gets$ frag(CS)\;
    \ForEach{$podsToReorder \in P_E$ \textbf{where} $|podsToReorder| < M_{ER}$}{
        CS' $\gets$ CS - \{podsToReorder on edge\}\;
        mapping, \_ $\gets$ match(CS', podsToReorder)\;
        CS' $\gets$ CS' $\cup$ \{deploying $podsToReorder$ by $mapping$\}\;
        currentFrag $\gets$ frag(CS')\;
        \If{currentFrag $<$ leastFrag}{
            leastFrag $\gets$ currentFrag\;
            bestMigrations $\gets$ \{(pod $\to$ mapping[pod]) for pod in $podsToReorder$\}\;
        }
    }
    \Return{bestMigrations}\;
}
\end{algorithm}

\textbf{appleDecision:} A decision may contain multiple components, and the applyDecision function (Algorithm \ref{alg:apply_decision}) manages each of these accordingly. First, it may include a set of migrations, which could involve migrating pods from cloud to edge, from edge to cloud, or reordering within the edge itself. The applyDecision function generates a multi-step plan for each migration and forwards it to the placement manager. In addition, there could be a list of new pods designated for deployment on the edge, each with a specific target node. For these pods, the function creates a single-step plan to bind them to their respective edge nodes. Lastly, for pods that need to be placed on the cloud, a single-step plan is created for each to be bound to the cloud node.
After selecting the best decision, the next step is to apply it. The algorithm begins by assigning each pod, intended for deployment on the edge, to an appropriate edge node, using the match algorithm based on the cluster state after the migrations. This results in a mapping of each pod to its target edge node. However, there may be cases where not all the pods intended for edge deployment can be successfully matched to a node, as the best decision was initially derived based on approximate feasibility. In such cases, as many pods as possible are deployed on the edge, while the remaining pods are redirected to the cloud.
\\
The placement manager then executes each plan, step by step. For any given plan, the scheduler ensures that no action begins until all preceding actions have been successfully completed. With this in mind, as detailed in algorithm \ref{alg:apply_decision}, plans for deploying pods on the cloud are independent and can proceed without inter-dependencies. However, the edge deployment plan is conditional upon the successful completion of all migrations (with an additional dependency created to ensure proper pod deployment on the edge). If a plan fails partway through, any remaining pods within that plan are deployed on the cloud, where they await further reordering by the suggester to move them to the edge.

\begin{algorithm}
\caption{Apply Decision Algorithm}\label{alg:apply_decision}
\KwIn{Cluster State, decision}

\SetKwFunction{FapplyDecision}{applyDecision}
\SetKwProg{Fn}{Function}{:}{}
\Fn{\FapplyDecision{CS, decision}}{
    CS' $\gets$ CS after migrating decision.migrations\;
    matching, remainingPods $\gets$ match(CS', decision.toEdgePods)\;

    toCloudPlans $\gets$ []\;
    \ForEach{pod in decision.toCloud $\cup$ remainingPods}{
        toCloudPlans $\gets$ append(toCloudPlans, [deploy pod on cloud])\;
    }

    toEdgePlan $\gets$ []\;
    \ForEach{migration in decision.migrations}{
        toEdgePlan $\gets$ (toEdgePlan..., then migration)\;
    }
    \ForEach{pod in matching}{
        toEdgePlan $\gets$ (toEdgePlan..., then deploy pod on edge)\;
    }

    \ForEach{plan in toCloudPlans}{
        execute plan\;
    }
    execute toEdgePlan\;
}
\end{algorithm}

\textbf{match:} The goal of the Algorithm \ref{alg:match} is to match new pods to edge nodes while satisfying the following conditions: maximize the number of pods matched, and prioritize matchings that result in the least amount of fragmentation among different options. To achieve this, the \verb|match| algorithm employs a dynamic programming approach.
Let $dp[(i, pods)]$ represent the minimum fragmentation possible when deploying a set of pods to edge nodes from $1$ to $i$. Based on this definition, for each $i$ from $1$ to the number of edge nodes $|E|$, and for each subset of new \verb|pods| $pods$, a state is defined for calculating $dp$.
The base state, representing the initial state of the cluster, is:
\begin{equation}
    dp[(0,\emptyset)]=frag(CS)
\end{equation}
All other states are initially marked as impossible (i.e. setting $dp$ value to $\infty$).
The recurrence relation is formulated as follows: for each state $(i, pods)$, the goal is to deploy a (possibly empty) subset of $pods$ to the $i$th node. We refer to this subset as $cur_{pods}$. First, it must be verified if $cur_{pods}$ can be deployed on the $i$th node, which is determined by comparing the remaining resources of the node with the total resources required by $cur_{pods}$. If the deployment is feasible, a possible solution for $dp[(i, pods)]$ can be derived as:
\begin{equation}
    \begin{aligned}
        dp[(i, pods)] & = \\
        & dp[(i - 1, pods - cur_{pods})] \\
        & + frag(E_i \cup cur_{pods}) - frag(E_i)
    \end{aligned}
\end{equation}
Based on this, the recurrence relation for $dp$ can be expressed as:
\begin{equation}
    \begin{aligned}
        dp[(i, pods)] & = \\
        & min_{\forall cur_{pods} \subseteq pods} \Big( dp[(i - 1, pods - cur_{pods})] + \\ 
        & frag(E_i - cur_{pods}) - frag(E_i) \Big)
    \end{aligned}
\end{equation}
The final solution is obtained by finding the largest subset of input pods (denoted as $final_{pods}$ such that $dp[(|E|, final_{pods})]$ is not $\infty$ and has the minimum fragmentation.
To determine the specific matching, the algorithm keeps track of dynamic programming updates in an additional array called $par$. For each state $S$, $par[S]$ stores $S'$, which is the state from which $dp[S]$ was updated. By tracing back from $par[final_{state}]$ to the base state $(0, \emptyset)$, the deployment of each pod to its respective node can be determined.

\begin{algorithm}
\caption{Match Algorithm}\label{alg:match}
\KwIn{Cluster State, newPods}
\KwOut{targets, the minimum fragmentation achieved}
\SetKwFunction{Fmatch}{match}
\SetKwProg{Fn}{Function}{:}{}
\Fn{\Fmatch{CS, newPods}}{
    dp $\gets$ mapping from all states $(i, pod_{set})$ to $\infty$\;
    par $\gets$ mapping from all states to None\;
    dp$(0, \emptyset) \gets$ frag(CS)\;
    par$(0, \emptyset) \gets$ Nil\;
    \ForEach{i $\gets$ 1 \textbf{to} $|E|$}{
        \ForEach{pods $\subseteq$ newPods}{
            \ForEach{$cur_{pods}$ $\subseteq$ $pods$}{
                fragChange $\gets$ frag(CS[$E_i$] $\cup$ $cur_{pods}$) $-$ frag(CS[$E_i$])\;
                
                achievedFragmentation $\gets$ dp$(i-1, pods -cur_{pods}) + $ fragChange\;
                \If{dp$(i, pods) > $ achievedFrag}{
                    dp$(i, pods) \gets$ achievedFrag\;
                    par$(i, pods) \gets (i-1, pods - cur_{pods})$\;
                }
            }
        }
    }
    max\_sub\_set $\gets \emptyset$\;
    \ForEach{pods $\subseteq$ newPods}{
        \If{dp$(|E|, pods) = \infty$}{
            \textbf{continue}\;
        }
        \If{$|\text{pods}| > |\text{max\_sub\_set}|$ \textbf{or} ($|\text{pods}| = |\text{max\_sub\_set}|$ \textbf{and} dp$(|E|, \text{pods}) < $ dp$(|E|, \text{max\_sub\_set}))$}{
            max\_sub\_set $\gets$ pods;
        }
    }
    targets $\gets \{\}$\;
    stateIterator $\gets (|E|, \text{max\_sub\_set})$\;
    \While{stateIterator $\neq$ None}{
        $i, \text{pods} \gets$ stateIterator\;
        $cur_{pods} \gets$ pods $-$ par$(i, \text{pods})$\;
        \ForEach{pod $\in$ $cur_{pods}$}{
            targets[pod] $\gets$ $E_i$\;
        }
        stateIterator $\gets$ par$(i, \text{pods})$\;
    }
    \ForEach{pod $\in$ newPods $-$ max\_sub\_set}{
        targets[pod] $\gets$ cloud node\;
    }
    \Return{targets, dp$(|E|, \text{max\_sub\_set})$}\;
}
\end{algorithm}

\subsection{Complexity analysis}
Let's analyze each algorithm in the reverse dependency order.

\textbf{match:} The matching algorithm has two main steps. First, it calculates the $dp$ value for all $|E|\times 2^{|newPods|}$ states, which is $O(|E|\times \sum_{N'\subset newPods}{2^{|N'|}})$ that is equal to $O(|E|\times 3^{|newPods|})$. Second, it uses these values to find the best match. This second step takes at most $|E|$ iterations to track the best solution and extract targets for each new pod, so it is $O(|E|+|newPods|)$. Overall the complexity is $O(|E|\times 3^{|newPods|})$.

\textbf{applyDecision:} This algorithm first uses the \verb|match| algorithm to fit $decision.toEdgePods$ to the edge, this parts will be done in $O(|E|\times 3^{|decision.toEdgePods|})$ complexity, and then will simply create a plan for each pod in the decision which is done in $|decision|$, in the end, the complexity is $O(|E|\times 3^{|decision.toEdgePods|} + |decision|)$.

\textbf{reorderEdge:} The algorithm invokes the \verb|match| function for each subset of $P_E$ that has at most $M_{ER}$ pods in it. So the complexity is always less than $O(|P_E|^{M_{ER}}\times |E| \times 3^{M_{ER}})$.

\textbf{freeEdgeAsNeeded:} In this algorithm, all edge pods are sorted in each iteration by a score function, and the best one is chosen. The scores can be computed at first for each deployment and modified whenever a pod from that service is chosen. With this implementation, the time complexity will be $O(|P_E|^2\times log(|P_E|))$.

\textbf{bestFit:} This algorithm simply calls two \verb|freeEdgeAsNeeded| and \verb|reorderEdge| functions in order. So the complexity is $O(|P_E|^{M_{ER}}\times |E| \times 3^{M_{ER}}+|P_E|^2\times log(|P_E|))$.

\textbf{makeDecision:} For each subset of new pods, if the \verb|doMigrate| is enabled, it will invoke the \verb|bestFit| function, otherwise, it will compare the resources. In both cases, it will calculate the QoS after and before the decision that can be implemented efficiently as described in \verb|freeEdgeAsNeeded|.This means, if \verb|doMigrater| is enabled then the complexity is $O(2^{|newPods|}\times Complexity_{bestFit})$, and otherwise $O(2^{|newPods|}\times |newPods|)$.

\textbf{suggest:} This is one of the main entry points of the scheduler. It consists of three parts, first choosing the cloud pods to suggest, which are the same as \verb|freeEdgeAsNeeded| pod choosing phase but it is for cloud pods, so the complexity is $O(M_{C2E}\times |P_C|\times log(|P_C|))$. The second part is calling \verb|makeDecision| on the chosen pods, with migration being enabled, which is done in $O(2^{M_{C2E}}\times (|P_E|^{M_{ER}}\times |E|\times 3^{M_{ER}}+|P_E|^2\times log(|P_E|)))$. The third part is applying the decision made for the suggested pods with $O(|E|\times 3^{M_{C2E}}+M_{C2E}+|P_E|)$ complexity. The $M_{C2E}+|P_E|$ term accounts for the maximum size of a decision (may migrate all $|P_E|$ edge pods to the cloud and migrate $M_{C2E}$ pods from the cloud to the edge). Overall, the complexity is: 
\begin{equation}
\label{eq:suggest-complexity}
    \begin{aligned}
    O ( & M_{C2E}\times |P_C|\times log(|P_C|) +\\
    & 2^{M_{C2E}}\times |P_E|^{M_{ER}}\times |E|\times 3^{M_{ER}} +\\
    & 2^{M_{C2E}}\times |P_E|^2\times log(|P_E|) \\
    & |E|\times 3^{M_{C2E}}
    )
    \end{aligned}
\end{equation}

\textbf{immediateSchedule:} This is the other main entry point of the scheduler. It includes two steps, first invoking \verb|makeDecision| on the $newPods$, with migration being disabled, which is done in $O(2^{|N|}\times |N|)$, and then apply the decision made using \verb|applyDecision| that is done in $O(|E|\times 3^{M_{C2E}}+M_{C2E}+|P_E|)$. Given this, the overall complexity is:
\begin{equation}
\label{eq:immediateSchedule-complexity}
    \begin{aligned}
    O ( & 2^{|N|}\times |N| +\\
    & |E|\times 3^{M_{C2E}}+M_{C2E}+|P_E|
    )
    \end{aligned}
\end{equation}

For further analysis, let's evaluate how large $M_{C2E}$ and $M_{ER}$ can get. The migration is not a built-in feature in Kubernetes and to simulate a migration, we have to delete the pod where it is while creating the new one. Because Kubernetes ensures having the HPA set number of pods always running, it won't terminate the pod, until the new pod is running. This may cause some migrations to take a two-step approach (i.e. migrate the pod first to the cloud, and then migrate it from the cloud to the target edge node) when the resources allocated to the pod are in need (for other migrations and scheduling). This entire process can take a couple of seconds to fulfill. Have this in mind, $M_{ER}$ is set from 1 to 3, meaning the scheduler can do at most three migrations per suggestion event (that happens periodically), where the source of the migration is in the edge. Because migrations from the cloud is easier (the cloud resources are infinite) $M_{C2E}$ can be set to larger numbers up to ten. The \verb|suggest| algorithm is linear to the number of pods on cloud $|P_C|$ and as the number of pods on the edge $|P_E|$ gets lower, it can search for more migrations on edge. The \verb|immediateSchedule| algorithm performance is mostly influenced by the number of new pods $|N|$. With this complexity we can schedule a batch of 20 pods simultaneously in less than a second, which is a great improvement over Kubernetes default or other greedy schedulers' ability to schedule one pod at a time.
\section{Evaluation}
\label{section:evaluation}
KubeDSM is implemented in Golang \cite{go_website} and operates as a pod within the cluster. It does not replace the Kube-scheduler pod; rather, it coexists with it. This design enables workloads to specify their desired scheduler by utilizing the "schedulerName" attribute in Kubernetes \cite{noauthor_kubernetes_nodate}. To establish connectivity with the API gateway, we utilized the Kubernetes Golang SDK \cite{kube_go_sdk_github} and configured the necessary cluster role bindings (RBACs) to enable the scheduler to allocate pods to nodes. In addition, we implemented another scheduler,  Sencillo\footnote{Sencillo translates to "simple" in Spanish}, which provides baseline algorithms for comparative analysis of KubeDSM's performance. The Project Sencillo baselines include Biggest Edge Node First (BEF), Smallest Edge Node First (SEF), and Cloud First (CF). We also compared the results against the default kube-scheduler (K8S).\\
\subsection{Cluster Setup}
\begin{table}[h]
\centering
\caption{Summary of Nodes configuration used in our evaluations}
\label{table:node_configuration}
\begin{tabular}{@{\extracolsep\fill}|c|c|c|c|}
\hline
   Node Name& Role                       & Memory (GB)& Cores\\ \hline
N1 & Master                     & 6      & 4           \\ \hline
N2 & \multirow{3}{*}{Edge Node} & 5      & 5           \\  \cmidrule{1-1} \cmidrule{3-4}
N3 &                            & 4      & 4           \\  \cmidrule{1-1} \cmidrule{3-4}
N4 &                            & 5      & 7           \\ \hline
N5 & Cloud Node                 & 17     & 22          \\ \hline
\end{tabular}
\end{table}

As detailed in Table \ref{table:node_configuration}, we established a K3S cluster \cite{noauthor_k3s_nodate}, consisting of one master node, three edge nodes, and one resource-rich cloud node. The cloud node is provisioned with more resources, while the edge nodes are equipped with heterogeneous resources to replicate typical configurations found in cloud-assisted edge clusters. We developed a custom load generation tool, DrStress \cite{noauthor_drstress_nodate}, to conduct load tests by simulating multi-threaded HTTP requests to our pods. Additionally, we configured Kubernetes's Horizontal Pod Autoscaler (HPA) to dynamically adjust the pod count based on incoming request rates. \\
\begin{table}[h]
\centering
\caption{Summary of service deployments' resource requests}
\label{table:deployment_configuration}
\begin{tabular}{|c|c|c|}
\hline
Service Name & Cores  & Memory (Mi) \\ \hline
A               & 1     & 950           \\ \hline
B               & 1     & 1900           \\ \hline
C               & 1     & 950           \\ \hline
D               & 2    & 1900           \\ \hline
\end{tabular}
\end{table}
Furthermore, we assumed four deployments in our cluster: A, B, C, and D, with their resource configurations detailed in Table \ref{table:deployment_configuration}. Each scenario initiates with one pod from each service deployment. As the request rate increases, HPA dynamically scales the number of pods, allowing us to evaluate the scheduler's performance across different conditions.
\subsection{Scenario Design}
We designed several workload scenarios, which DrStress uses to drive tests. These scenarios generate request rates based on the normal distribution to evaluate the schedulers' efficiency. Each scenario features different request rates, demanding varying amounts of cluster resources. For clarity, scenarios are named according to the amount of edge resources required. For example, a scenario labeled 1.0\_0.5 signifies a generated workload that fully utilizes edge resources on average (100\%), with a standard deviation of 0.5 in the normal distribution.

Each scenario consists of 12 cycles, each lasting 90 seconds. For each cycle, the amount of resources used is determined by sampling the usage fraction from the distribution. These resources are then equally divided among deployments. Subsequently, for each cycle, the request rate is calculated to trigger the Horizontal Pod Autoscaler (HPA) to create the desired number of pods for each deployment. These calculated request rates are then passed to DrStress to execute the scenario.

Since the default kube-scheduler does not distinguish between cloud and edge nodes, directly comparing it with KubeDSM and other edge-oriented schedulers would be inequitable. To mitigate this, we employed Kubernetes's "Taint \& Toleration" feature, which allows nodes to repel a set of pods, thereby prioritizing edge nodes for pod allocation \cite{taint_toleration_kube}. In addition, we set node affinity for the pods to attract them to the edge nodes.

\subsection{Results}
\begin{figure}[ht]
    \centering
        \includegraphics[width=0.95\linewidth]{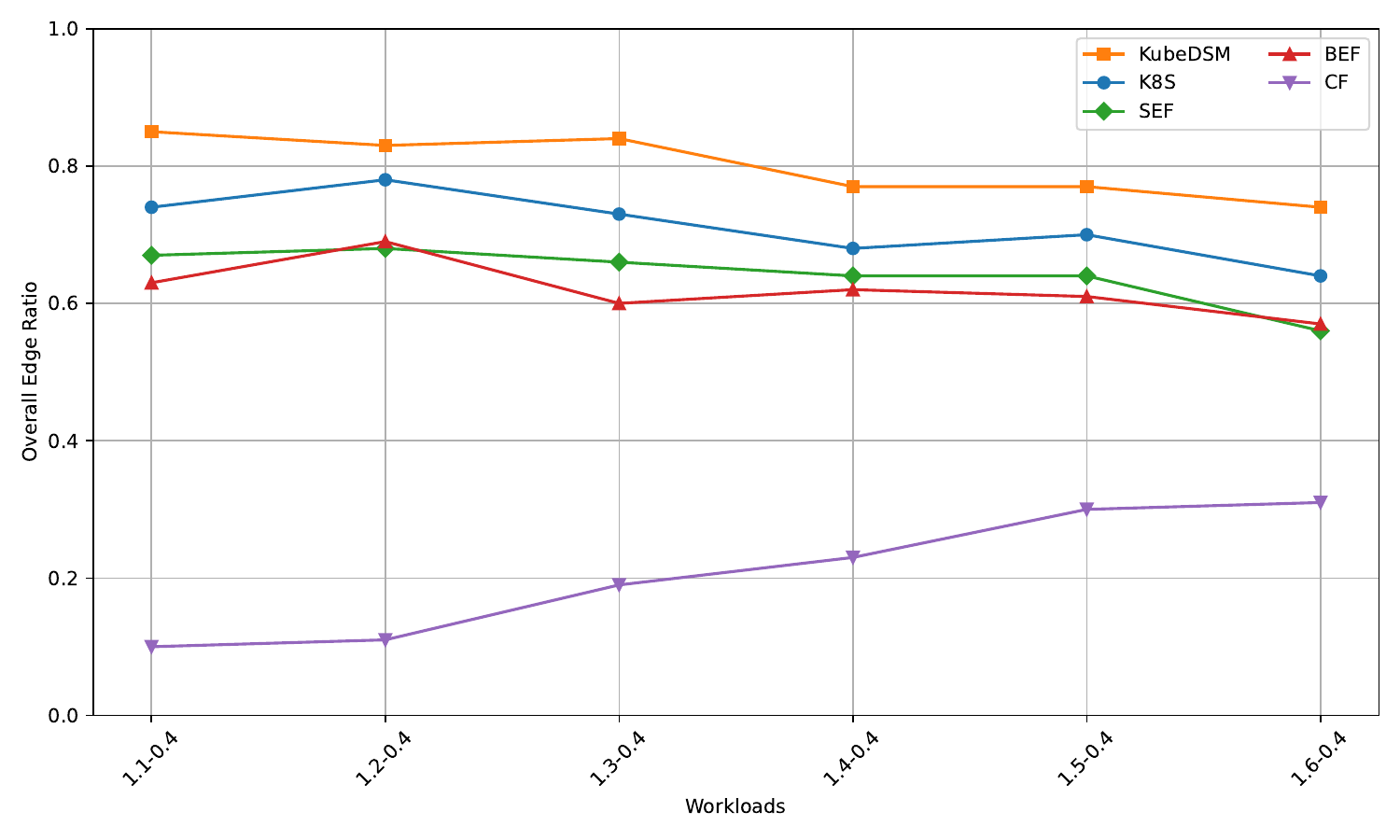}
        \caption{Comparison of edge ratio under varying average request rates.}
        \label{fig:trend1_edge_ratio_overall_line}
\end{figure}
    \begin{figure}[ht]
    \centering
        \includegraphics[width=0.95\linewidth]{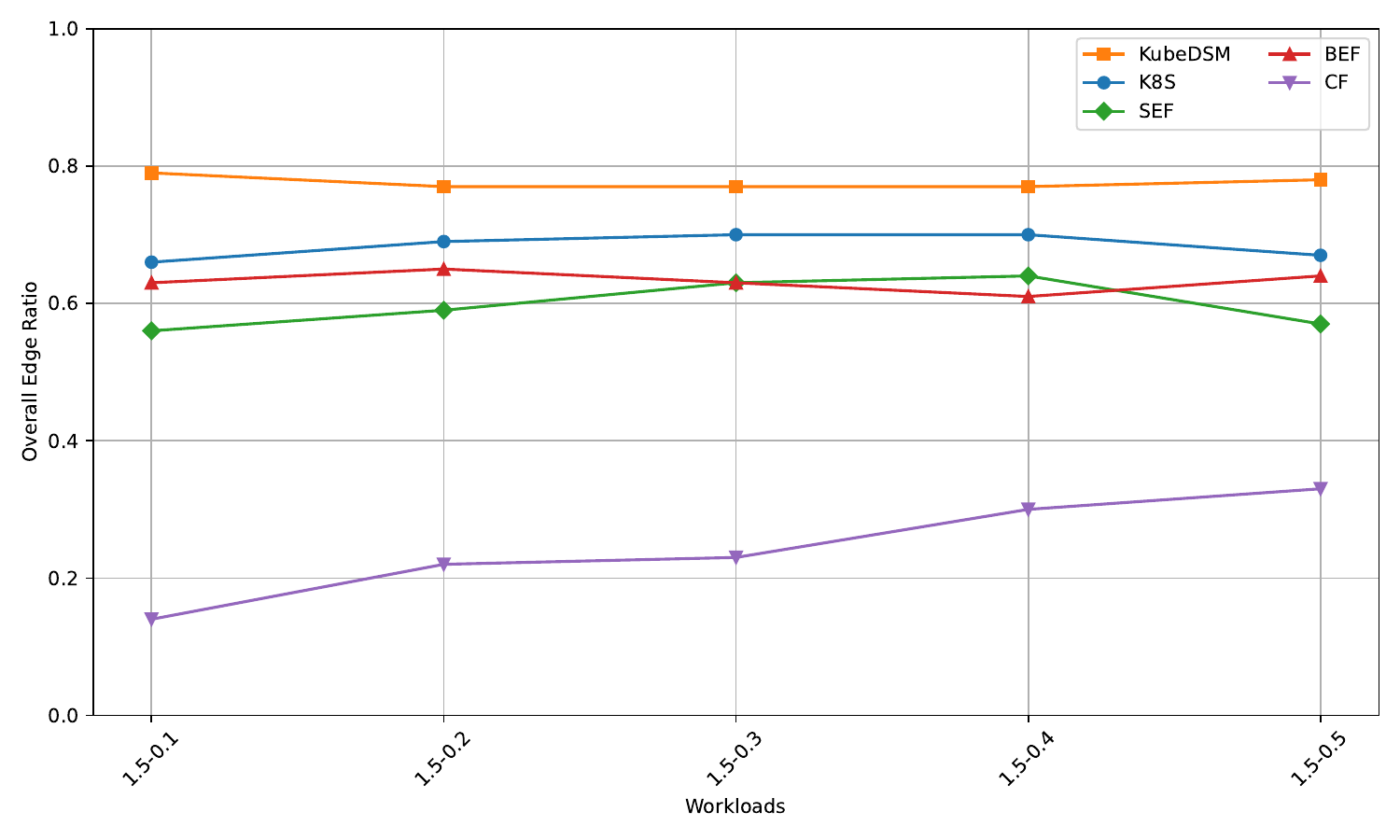}
        \caption{Comparison of edge ratio under different amount of variation in request rates.}
        \label{fig:trend2_edge_ratio_overall_line}
\end{figure}

In the first set of experiments, we evaluated the performance of KubeDSM by examining the edge ratio, under various workload scenarios. This metric is defined as the average percentage of pods scheduled to edge nodes across all deployments. Figure~\ref{fig:trend1_edge_ratio_overall_line} demonstrates the comparison of edge ratios as the average request rates increase from 1.1 to 1.6, with the standard deviation fixed at 0.4. The results indicate that KubeDSM consistently surpasses other algorithms in terms of average edge ratio, highlighting its superior capability to schedule a larger number of pods to edge nodes. The default kube-scheduler (K8S), SEF, and BEF exhibit similar performance levels, while CF performs the worst. On average, KubeDSM achieved an edge ratio of 80\%, representing an improvement of 13.4\% compared to the best-performing baseline across all workloads. Figure~\ref{fig:trend2_edge_ratio_overall_line} illustrates the comparison of edge ratios as the standard deviation of request rates increases from 0.1 to 0.5, while keeping the average fixed at 1.5. The results exhibit similar trends, demonstrating an average edge ratio of 78\% for KubeDSM, with an overall improvement of 18.9\% compared to the best-performing baseline across all workloads.\\

To better illustrate how effectively the schedulers increase the edge ratio for each deployment, we present a comparison of edge ratios across different workloads for each individual deployment separately. 
\begin{figure}[ht]
    \centering
\includegraphics[width=0.95\linewidth]{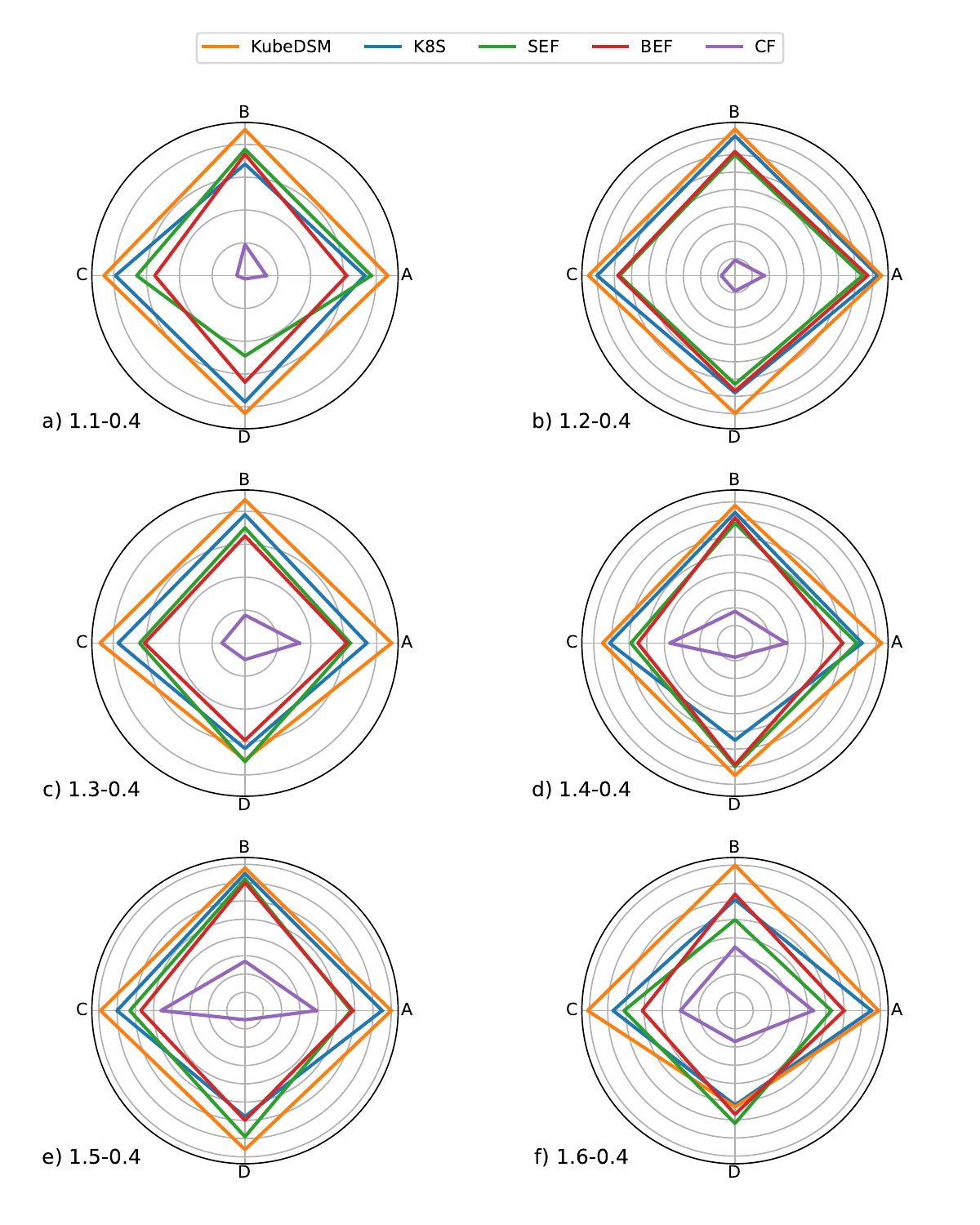}
        \caption{Comparison of edge ratio for each individual deployment under varying average request rates.}
        \label{fig:trend1_radar}
\end{figure}
    \begin{figure}[ht]
    \centering
\includegraphics[width=0.95\linewidth]{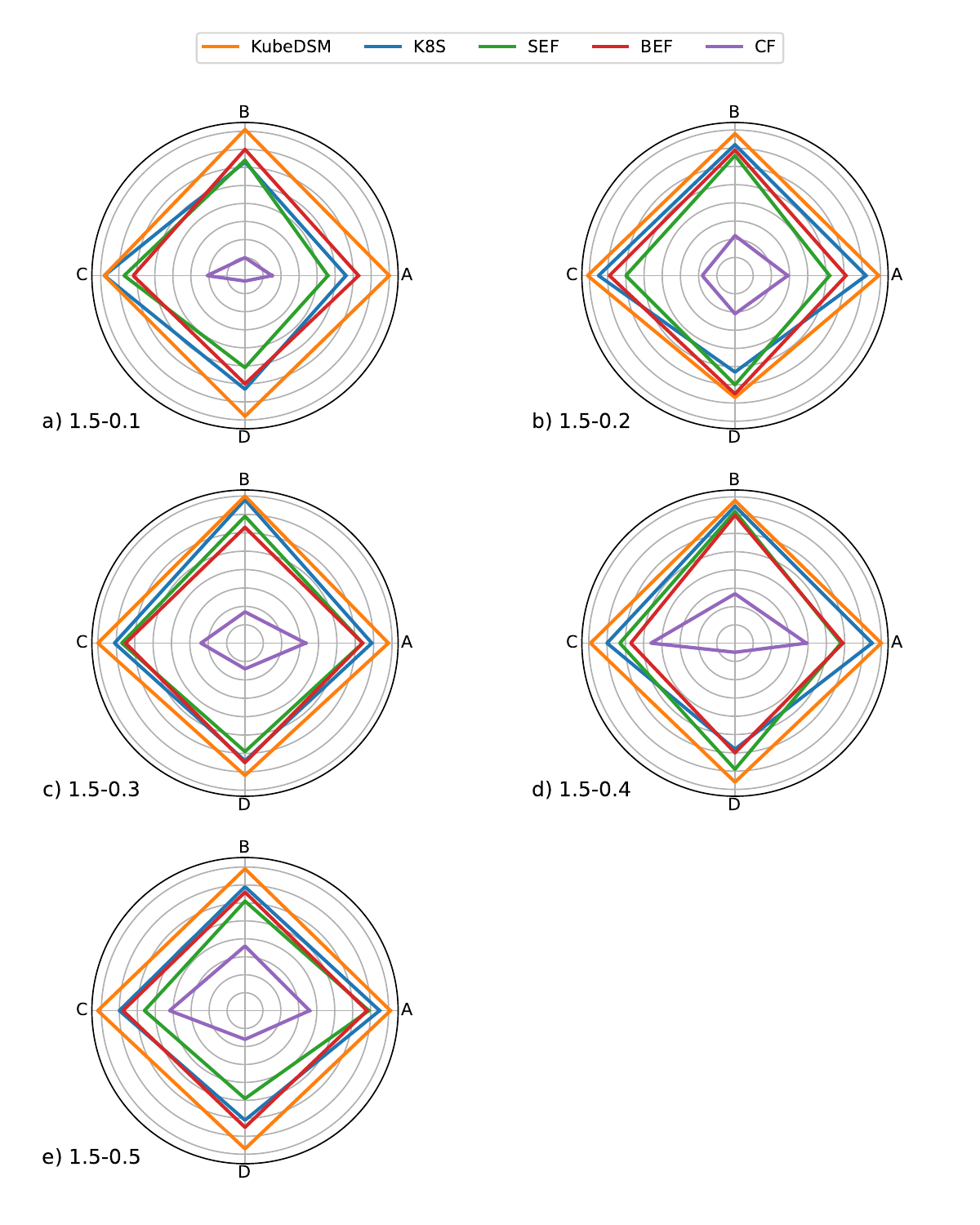}
        \caption{Comparison of edge ratio for each individual deployment under different amount of variation in request rates.}
        \label{fig:trend2_radar}
\end{figure}

\begin{figure}[ht]
    \centering
\includegraphics[width=0.95\linewidth]{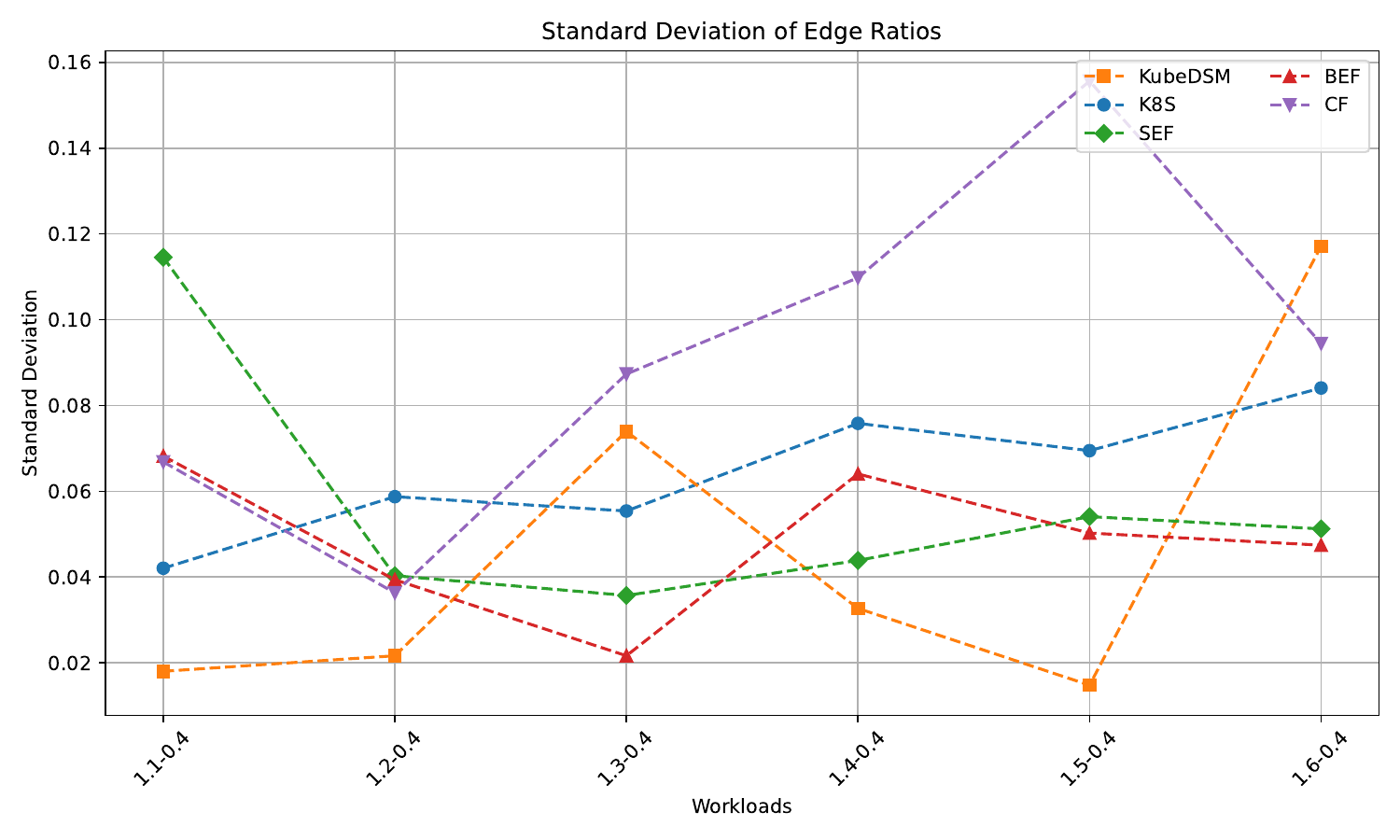}
        \caption{Comparison of edge ratio standard deviation across all deployments under varying average request rates.}
        \label{fig:trend1_std_deviation}
\end{figure}

\begin{figure}[ht]
    \centering
\includegraphics[width=0.95\linewidth]{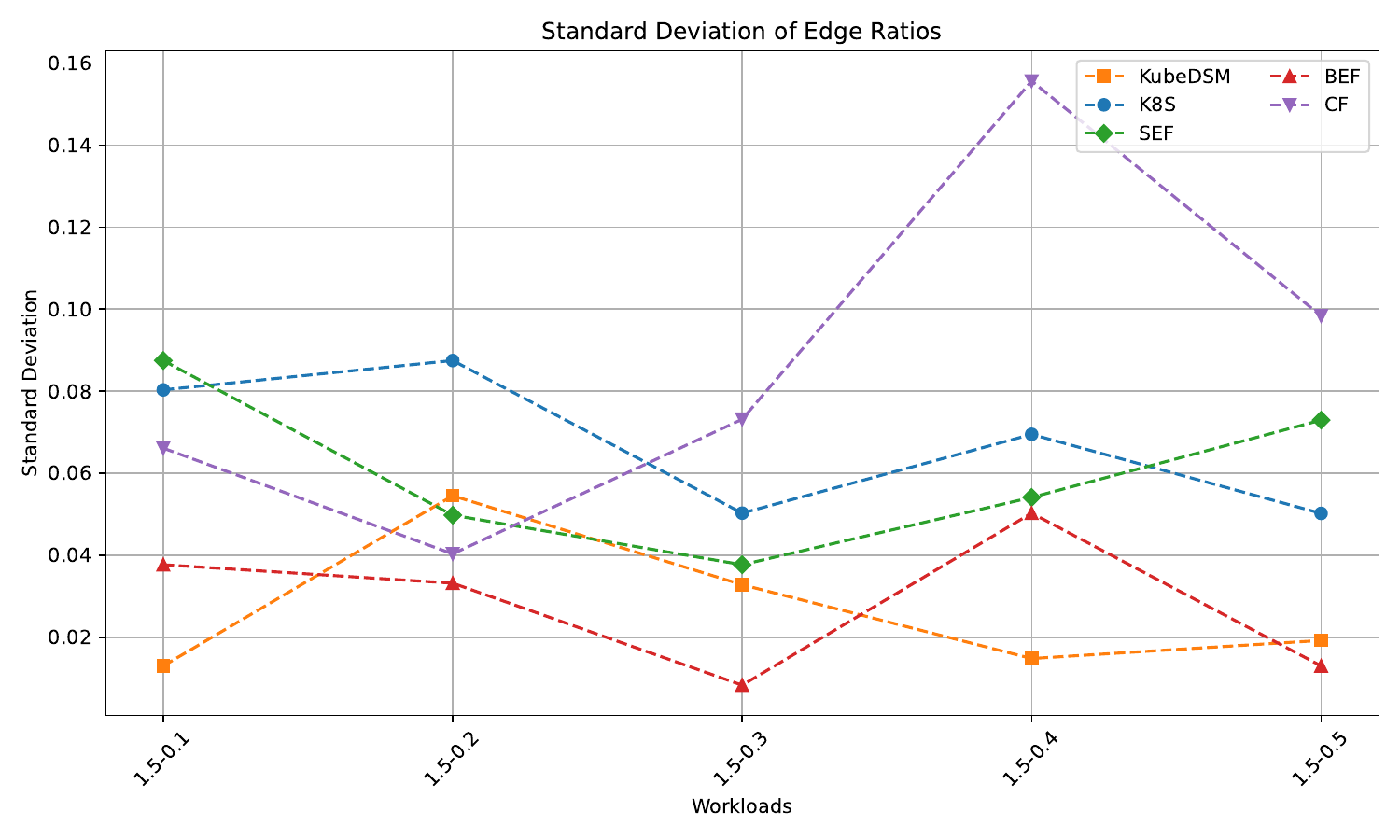}
        \caption{Comparison of edge ratio standard deviation across all deployments under varying average request rates.}
        \label{fig:trend2_std_deviation}
\end{figure}

Figure~\ref{fig:trend1_radar} illustrates the comparison of edge ratios for each individual deployment as the average request rates increase, as shown in sub-figures (a-f). In each sub-figure, the radar chart radius represents a scale from 0 to 100\% edge ratio. Therefore, the closer the vertex for each deployment is to the outer edge of the chart, the better the edge ratio for that deployment. Additionally, the more symmetrical the radar chart is, the better the scheduler has balanced the deployments. As also supported by Figures~\ref{fig:trend1_edge_ratio_overall_line} and \ref{fig:trend2_edge_ratio_overall_line}, KubeDSM consistently achieves a superior edge ratio compared to the other baselines. Additionally, it demonstrates a better balance between deployments. However, an interesting observation is that as the load increases to 1.6, KubeDSM tends to sacrifice Deployment D in favor of other deployments. Since Deployment D is larger than the others, it prioritized the smaller deployments in order to achieve higher overall edge ratio. Figure~\ref{fig:trend2_radar} also illustrates the comparison of edge ratios for each individual deployment as the variation in request rates increase. The results reveals that KubeDSM achieves a better balance between deployments in these experiments as well. To quantify the balance, we present the standard deviation of edge ratios across all deployments in Figures~\ref{fig:trend1_std_deviation}-\ref{fig:trend2_std_deviation}. The results indicate that KubeDSM achieved a lower standard deviation in edge ratios across 6 out of 11 workloads among all other algorithms. Compared to K8S which is the second best performing algorithm in terms of average edge ratio, KubeDSM not only achieves better edge ratio in all 11 workload scenarios, but also obtains lower standard deviation in 9 out of 11 scenarios as well. Achieving a higher average with a lower variation indicates that KubeDSM not only schedules more pods to edge nodes effectively but also does so consistently across different deployments. 

In a separate set of experiments, we evaluated the impact of migrations on the edge ratios achieved by KubeDSM. KubeDSM features two configuration parameters, $M_{C2E}$ and $M_{ER}$, which regulate the number of migrations. By default (denoted as \textbf{Original} in the figures), these parameters are set to 5 and 3, respectively. We developed three different variations of KubeDSM by modifying the values of these parameters:

\textbf{MidMig}: With $M_{C2E}=2$ and $M_{ER}=1$, this configuration assesses the effect of reducing the number of migrations.

\textbf{NoCloudOffload}: Setting $M_{C2E}=0$ and $M_{ER}=3$, this variation evaluates the impact of disabling cloud-to-edge migrations.

\textbf{NoMig}: With $M_{C2E}=5$ and $M_{ER}=0$, this setup examines the effect of disabling edge reordering.

\begin{figure}[ht]
    \centering
\includegraphics[width=0.95\linewidth]{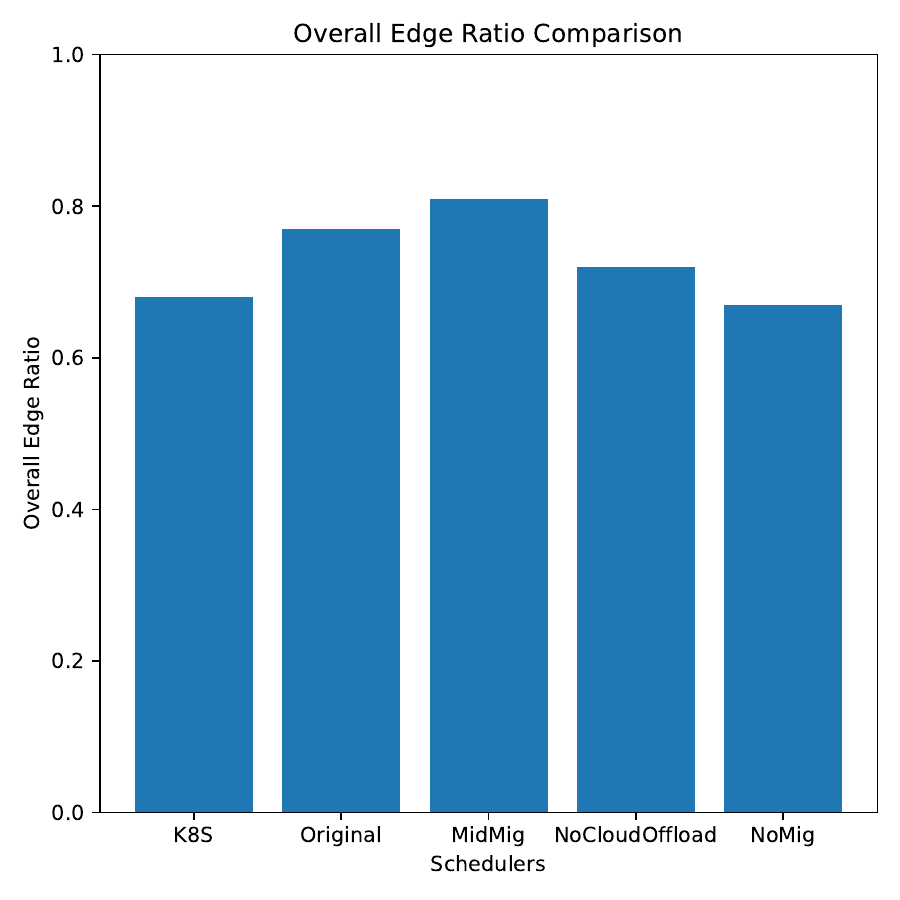}
        \caption{Comparison of edge ratio across different migration scenarios}
        \label{fig:migration}
\end{figure}

Figure~\ref{fig:migration} presents a comparison of the average edge ratios between the default K8S scheduler, KubeDSM, and its variations. NoMig shows a higher edge ratio compared to K8S, demonstrating that KubeDSM can improve the edge ratio by an average of 8\% solely through its scheduling strategy, even without any migrations. There is up to a 20.8\% improvement between the best and worst KubeDSM variations, confirming that migration significantly enhances the effectiveness of the method. NoCloudOffload exhibits a better edge ratio compared to NoMig, verifying the importance of edge reordering. On average, Original demonstrates 14.9\% improvement in terms of edge ratio over NoMig. Among all variations, MidMig exhibited the best results, surpassing even the Original configuration by 5.1\%. This finding supports the notion that enabling migration on edge clusters can enhance the edge ratio by reducing fragmentation. However, it also suggests that excessively increasing the number of migrations could have adverse effects.

\begin{figure}[ht]
    \centering
\includegraphics[width=0.7\linewidth]{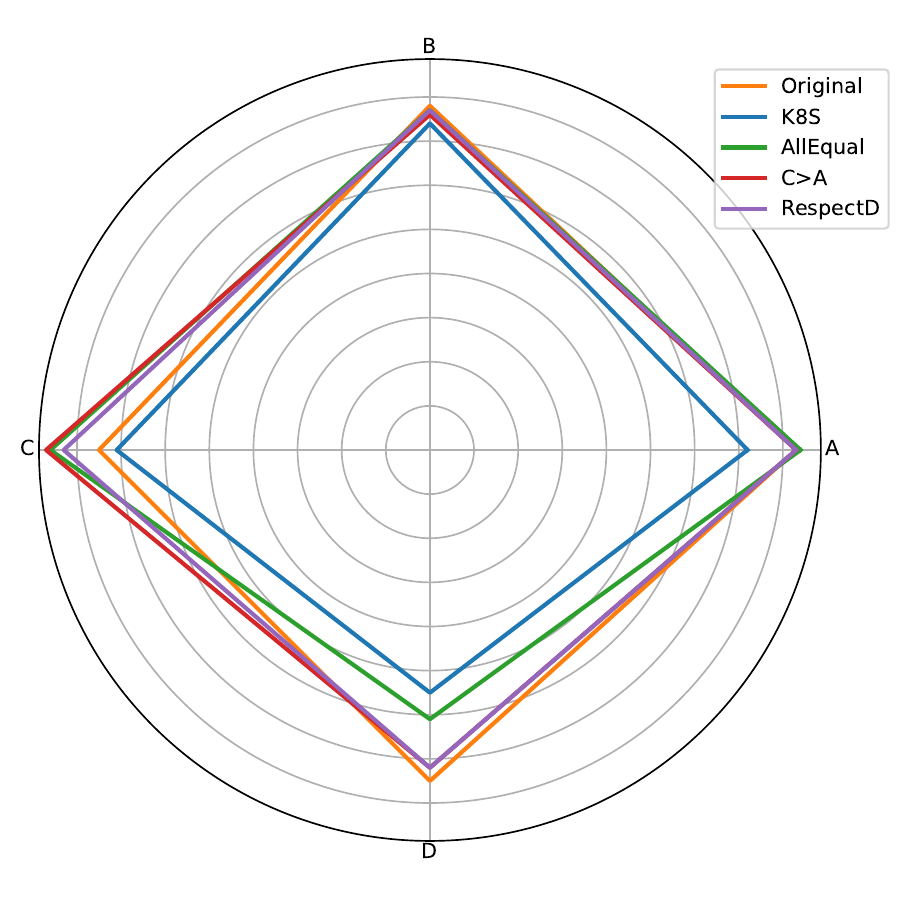}
        \caption{Comparison of edge ratio across different QoS scenarios}
        \label{fig:qos}
\end{figure}

In the final set of experiments, we evaluated the impact of different QoS configurations on the edge ratios achieved by KubeDSM. By default, KubeDSM sets the QoS target for all deployments to 1, meaning a deployment meets the QoS only if all of its pods are assigned to the edge. To assess how changing the QoS target affects the edge ratio, we modified this parameter to create the following scenarios:

\textbf{AllEqual}: Setting QoS values of all deployments to 0.5

\textbf{C$>$A}: Setting QoS values to 0.5, 0.1, 1, 0.1 for deployments A, B, C, and D respectively, giving higher priority to C, then A, followed by B and D.

\textbf{RespectD}: Setting the QoS value of the deployment D to 0.5 and all other deployments to 0.1, to prioritize D over the others. 

Figure~\ref{fig:qos} presents the results of the comparison across different QoS scenarios. In the AllEqual scenario, the edge ratio of deployment D decreased, while the edge ratio of deployment C increased. This occurred because deployment D is larger than the other deployments, and after satisfying D by bringing half of its pods to the edge, KubeDSM focused on increasing the overall edge ratio by allocating more pods to the edge from the smaller deployments. In the C$>$A scenario, KubeDSM increased the edge ratio of deployment C over the other deployments, as expected. Finally, in the RespectD scenario, there was a slight improvement in the edge ratio of deployment D, also as expected. These results confirm that KubeDSM aims to respect the QoS requirements of deployments while maximizing the overall edge ratio of all deployments.

\section{Conclusion}
\label{section:conclusion}
In this study, we evaluated the effectiveness of KubeDSM in comparison to the default kube-scheduler (K8S) and other baseline variations. Our results demonstrate that KubeDSM consistently achieves a higher average edge ratio and a lower standard deviation in edge ratios, indicating not only a more effective scheduling strategy but also greater consistency and stability across various deployments. 

Through detailed experiments, we observed that enabling migrations on edge clusters significantly enhances the overall edge ratio by reducing fragmentation. The MidMig variation, in particular, yielded the best results, even outperforming the original configuration, suggesting that while enabling migrations can improve performance, an excessive number of migrations may have adverse effects.

Finally, our experiments on QoS configurations revealed that KubeDSM effectively respects the QoS requirements of deployments while striving to maximize the overall edge ratio. Different QoS scenarios, such as AllHalf, C$>$A, and RespectD, showed KubeDSM's adaptability in balancing the edge ratio based on the specified priorities and constraints.

Overall, the findings highlight KubeDSM's capability to optimize edge resource utilization and provide a balanced and efficient scheduling solution for edge computing environments. Future work could explore further enhancements to KubeDSM's migration strategies and QoS configurations to achieve even better performance and adaptability in diverse deployment scenarios.

\bibliography{references}

\begin{thebibliography}{10}
\expandafter\ifx\csname url\endcsname\relax
  \def\url#1{\burl{#1}}\fi
\expandafter\ifx\csname urlprefix\endcsname\relax\def\urlprefix{URL }\fi
\providecommand{\bibinfo}[2]{#2}
\providecommand{\eprint}[2][]{\url{#2}}
\providecommand{\doi}[1]{\url{https://doi.org/#1}}
\bibcommenthead

\bibitem{noauthor_google_nodate}
\bibinfo{title}{Google cloud | distributed cloud}.
\newblock \urlprefix\url{https://cloud.google.com/distributed-cloud}.

\bibitem{aws_amazon_nodate}
\bibinfo{title}{Amazon web services | local zones}.
\newblock
  \urlprefix\url{https://aws.amazon.com/about-aws/global-infrastructure/localzones/}.

\bibitem{azure_azure_nodate}
\bibinfo{title}{Azure private multi-access edge computing}.
\newblock
  \urlprefix\url{https://docs.microsoft.com/en-us/azure/private-multi-access-edge-compute-mec/overview}.

\bibitem{mohan_pruning_2020}
\bibinfo{author}{Mohan, N.} \emph{et~al.}
\newblock \emph{\bibinfo{title}{Pruning edge research with latency shears}},
  \bibinfo{pages}{182--189} (\bibinfo{publisher}{{ACM}}).
\newblock \urlprefix\url{https://dl.acm.org/doi/10.1145/3422604.3425943}.

\bibitem{ma_cost-efficient_2021}
\bibinfo{author}{Ma, X.} \emph{et~al.}
\newblock \bibinfo{title}{Cost-efficient resource provisioning for dynamic
  requests in cloud assisted mobile edge computing}
  \textbf{\bibinfo{volume}{9}}, \bibinfo{pages}{968--980}.
\newblock \urlprefix\url{https://ieeexplore.ieee.org/document/8660570/}.

\bibitem{ma_cost-efficient_2017}
\bibinfo{author}{Ma, X.} \emph{et~al.}
\newblock \emph{\bibinfo{title}{Cost-efficient workload scheduling in cloud
  assisted mobile edge computing}}, \bibinfo{pages}{1--10}
  (\bibinfo{publisher}{{IEEE}}).
\newblock \urlprefix\url{http://ieeexplore.ieee.org/document/7969148/}.

\bibitem{li_heterogeneity-aware_2020}
\bibinfo{author}{Li, C.}, \bibinfo{author}{Bai, J.}, \bibinfo{author}{Ge, Y.}
  \& \bibinfo{author}{Luo, Y.}
\newblock \bibinfo{title}{Heterogeneity-aware elastic provisioning in
  cloud-assisted edge computing systems} \textbf{\bibinfo{volume}{112}},
  \bibinfo{pages}{1106--1121}.
\newblock
  \urlprefix\url{https://linkinghub.elsevier.com/retrieve/pii/{S0167739X20300339}}.

\bibitem{wang_dynamic_2017}
\bibinfo{author}{Wang, S.} \emph{et~al.}
\newblock \bibinfo{title}{Dynamic service placement for mobile micro-clouds
  with predicted future costs} \textbf{\bibinfo{volume}{28}},
  \bibinfo{pages}{1002--1016}.
\newblock \urlprefix\url{http://ieeexplore.ieee.org/document/7557016/}.

\bibitem{kim_optimal_2021}
\bibinfo{author}{Kim, T.}, \bibinfo{author}{Al-Tarazi, M.},
  \bibinfo{author}{Lin, J.-W.} \& \bibinfo{author}{Choi, W.}
\newblock \bibinfo{title}{Optimal container migration for mobile edge
  computing: algorithm, system design and implementation}
  \textbf{\bibinfo{volume}{9}}, \bibinfo{pages}{158074--158090}.
\newblock \urlprefix\url{https://ieeexplore.ieee.org/document/9628116/}.

\bibitem{chen_service_2023}
\bibinfo{author}{Chen, W.}, \bibinfo{author}{Chen, Y.} \& \bibinfo{author}{Liu,
  J.}
\newblock \bibinfo{title}{Service migration for mobile edge computing based on
  partially observable markov decision processes}
  \textbf{\bibinfo{volume}{106}}, \bibinfo{pages}{108552}.
\newblock
  \urlprefix\url{https://linkinghub.elsevier.com/retrieve/pii/S0045790622007674}.

\bibitem{li_re-scheduling_2023}
\bibinfo{author}{Li, X.} \emph{et~al.}
\newblock \bibinfo{title}{Re-scheduling {IoT} services in edge networks}
  \textbf{\bibinfo{volume}{20}}, \bibinfo{pages}{3233--3246}.
\newblock \urlprefix\url{https://ieeexplore.ieee.org/document/10039683/}.

\bibitem{chi_multi-criteria_2023}
\bibinfo{author}{Chi, H.~R.} \emph{et~al.}
\newblock \bibinfo{title}{Multi-criteria dynamic service migration for
  ultra-large-scale edge computing networks} \textbf{\bibinfo{volume}{19}},
  \bibinfo{pages}{11115--11127}.
\newblock \urlprefix\url{https://ieeexplore.ieee.org/document/10043024/}.

\bibitem{rong_live_2023}
\bibinfo{author}{Rong, C.}, \bibinfo{author}{Wang, J.~H.},
  \bibinfo{author}{Wang, J.}, \bibinfo{author}{Zhou, Y.} \&
  \bibinfo{author}{Zhang, J.}
\newblock \bibinfo{title}{Live migration of video analytics applications in
  edge computing} \bibinfo{pages}{1--15}.
\newblock \urlprefix\url{https://ieeexplore.ieee.org/document/10049158/}.

\bibitem{ghafouri_mobile-kube_2022}
\bibinfo{author}{Ghafouri, S.} \emph{et~al.}
\newblock \emph{\bibinfo{title}{Mobile-kube: Mobility-aware and
  energy-efficient service orchestration on kubernetes edge servers}},
  \bibinfo{pages}{82--91} (\bibinfo{publisher}{{IEEE}}).
\newblock \urlprefix\url{https://ieeexplore.ieee.org/document/10061810/}.

\bibitem{lai_delay-aware_2023}
\bibinfo{author}{Lai, W.-K.}, \bibinfo{author}{Wang, Y.-C.} \&
  \bibinfo{author}{Wei, S.-C.}
\newblock \bibinfo{title}{Delay-aware container scheduling in kubernetes}
  \textbf{\bibinfo{volume}{10}}, \bibinfo{pages}{11813--11824}.
\newblock \urlprefix\url{https://ieeexplore.ieee.org/document/10044213/}.

\bibitem{marchese_2022}
\bibinfo{author}{Marchese, A.} \& \bibinfo{author}{Tomarchio, O.}
\newblock \emph{\bibinfo{title}{Network-aware container placement in cloud-edge
  kubernetes clusters}}, \bibinfo{pages}{859--865} (\bibinfo{publisher}{IEEE},
  \bibinfo{year}{2022}).
\newblock \urlprefix\url{https://ieeexplore.ieee.org/document/9826099/}.

\bibitem{chiaro_2024}
\bibinfo{author}{Chiaro, C.}, \bibinfo{author}{Monaco, D.},
  \bibinfo{author}{Sacco, A.}, \bibinfo{author}{Casetti, C.} \&
  \bibinfo{author}{Marchetto, G.}
\newblock \emph{\bibinfo{title}{Latency-aware scheduling in the cloud-edge
  continuum}}, \bibinfo{pages}{1--5} (\bibinfo{publisher}{IEEE},
  \bibinfo{year}{2024}).
\newblock \urlprefix\url{https://ieeexplore.ieee.org/document/10575183/}.

\bibitem{rausch_2021}
\bibinfo{author}{Rausch, T.}, \bibinfo{author}{Rashed, A.} \&
  \bibinfo{author}{Dustdar, S.}
\newblock \bibinfo{title}{Optimized container scheduling for data-intensive
  serverless edge computing}.
\newblock \emph{\bibinfo{journal}{Future Generation Computer Systems}}
  \textbf{\bibinfo{volume}{114}}, \bibinfo{pages}{259--271}
  (\bibinfo{year}{2021}).
\newblock
  \urlprefix\url{https://linkinghub.elsevier.com/retrieve/pii/{S0167739X2030399X}}.

\bibitem{qiao_2024}
\bibinfo{author}{Qiao, Y.} \emph{et~al.}
\newblock \bibinfo{title}{{EdgeOptimizer}: A programmable containerized
  scheduler of time-critical tasks in kubernetes-based edge-cloud clusters}.
\newblock \emph{\bibinfo{journal}{Future Generation Computer Systems}}
  \textbf{\bibinfo{volume}{156}}, \bibinfo{pages}{221--230}
  (\bibinfo{year}{2024}).
\newblock
  \urlprefix\url{https://linkinghub.elsevier.com/retrieve/pii/{S0167739X24000748}}.

\bibitem{ding_2023}
\bibinfo{author}{Ding, Z.}, \bibinfo{author}{Wang, S.} \&
  \bibinfo{author}{Jiang, C.}
\newblock \bibinfo{title}{Kubernetes-oriented microservice placement with
  dynamic resource allocation}.
\newblock \emph{\bibinfo{journal}{{IEEE} Trans. Cloud Comput.}}
  \textbf{\bibinfo{volume}{11}}, \bibinfo{pages}{1777--1793}
  (\bibinfo{year}{2023}).
\newblock \urlprefix\url{https://ieeexplore.ieee.org/document/9741392/}.

\bibitem{zhang_effective_2023}
\bibinfo{author}{Zhang, Q.}, \bibinfo{author}{Li, C.}, \bibinfo{author}{Huang,
  Y.} \& \bibinfo{author}{Luo, Y.}
\newblock \bibinfo{title}{Effective multi-controller management and adaptive
  service deployment strategy in multi-access edge computing environment}
  \textbf{\bibinfo{volume}{138}}, \bibinfo{pages}{103020}.
\newblock
  \urlprefix\url{https://linkinghub.elsevier.com/retrieve/pii/S1570870522001925}.

\bibitem{noauthor_kube-scheudler_nodate}
\bibinfo{title}{Kube-scheudler}.
\newblock
  \urlprefix\url{https://kubernetes.io/docs/reference/command-line-tools-reference/kube-scheduler/}.

\bibitem{noauthor_kube-scheduler-eviction_nodate}
\bibinfo{title}{Kube-scheduler eviction}.
\newblock
  \urlprefix\url{https://kubernetes.io/docs/concepts/scheduling-eviction/kube-scheduler/}.

\bibitem{noauthor_prometheus_nodate}
\bibinfo{title}{Prometheus website}.
\newblock \urlprefix\url{https://prometheus.io/}.

\bibitem{noauthor_grafana_nodate}
\bibinfo{title}{Grafana website}.
\newblock \urlprefix\url{https://grafana.com/}.

\bibitem{go_website}
\bibinfo{title}{The go programming language}.
\newblock \urlprefix\url{https://go.dev/}.

\bibitem{noauthor_kubernetes_nodate}
\bibinfo{title}{Kubernetes documentations}.
\newblock
  \urlprefix\url{https://kubernetes.io/docs/tasks/extend-kubernetes/configure-multiple-schedulers/}.

\bibitem{kube_go_sdk_github}
\bibinfo{title}{Kubernetes golang {SDK}}.
\newblock \urlprefix\url{https://github.com/kubernetes/client-go}.

\bibitem{noauthor_k3s_nodate}
\bibinfo{title}{k3s | lightweight kubernetes}.
\newblock \urlprefix\url{https://k3s.io}.

\bibitem{noauthor_drstress_nodate}
\bibinfo{title}{{DrStress} github}.
\newblock \urlprefix\url{https://github.com/AUT-Cloud-Lab/drstress}.

\bibitem{taint_toleration_kube}
\bibinfo{title}{Taint and tolerations \textbar kubernetes}.
\newblock
  \urlprefix\url{https://kubernetes.io/docs/concepts/scheduling-eviction/taint-and-toleration/}.

\end{thebibliography}
\end{document}